\documentclass[journal]{IEEEtran}
\usepackage{blindtext}
\usepackage{graphicx}
\usepackage{mathtools}
\usepackage{amssymb}
\usepackage[colorlinks]{hyperref}
\usepackage{algorithm}
\usepackage[noadjust]{cite}
\usepackage{float}
\hypersetup{citecolor=blue,linkcolor=blue}
\usepackage[font=footnotesize]{caption}
\usepackage[labelformat=simple]{subcaption}

\graphicspath{{figs/}}
\usepackage{bm}

\newfont{\bbb}{msbm10 scaled 700}

\newcommand{\av}{{\bf a}}

\newcommand{\cv}{{\bf c}}

\newcommand{\ev}{{\bf e}}
\newcommand{\fv}{{\bf f}}

\newcommand{\hv}{{\bf h}}

\newcommand{\rv}{{\bf r}}

\newcommand{\wv}{{\bf w}}
\newcommand{\vv}{{\bf v}}
\newcommand{\xv}{{\bf x}}

\newcommand{\zerov}{{\bf 0}}

\newcommand{\Am}{{\bf A}}
\newcommand{\Bm}{{\bf B}}
\newcommand{\Cm}{{\bf C}}
\newcommand{\Dm}{{\bf D}}

\newcommand{\Fm}{{\bf F}}

\newcommand{\Hm}{{\bf H}}
\newcommand{\Id}{{\bf I}}

\newcommand{\Lm}{{\bf L}}

\newcommand{\Pm}{{\bf P}}
\newcommand{\Qm}{{\bf Q}}

\newcommand{\Sm}{{\bf S}}

\newcommand{\Um}{{\bf U}}

\newcommand{\Lc}{{\cal L}}
\newcommand{\Mc}{{\cal M}}
\newcommand{\Nc}{{\cal N}}
\newcommand{\Oc}{{\cal O}}
\newcommand{\Pc}{{\cal P}}

\newcommand{\Uc}{{\cal U}}

\newcommand{\muv}{\hbox{\boldmath$\mu$}}

\newcommand{\sigmav}{\hbox{\boldmath$\sigma$}}

\newcommand{\Sigmam}{\hbox{\boldmath$\Sigma$}}

\newcommand{\Thetam}{\hbox{\boldmath$\Theta$}}

\renewcommand{\det}{{\hbox{det}}}

\newcommand{\transp}{{\sf T}}

\DeclareMathOperator*{\argmin}{arg\,min}
\DeclareMathOperator*{\argmax}{arg\,max}
\DeclareMathOperator{\Tr}{Tr}
\DeclareMathOperator{\dg}{diag}

\begin{document}
%
\title{Pre-demosaic Graph-based Light Field Image Compression}
%
%
%

\author{Yung-Hsuan Chao,
        Haoran Hong,~\IEEEmembership{Student Member,~IEEE,}
        Gene Cheung,~\IEEEmembership{Fellow,~IEEE,}
        and~Antonio Ortega,~\IEEEmembership{Fellow,~IEEE}
\thanks{Yung-Hsuan Chao was with University of Southern California. She is now with Qualcomm Technologies, Inc., 5775 Morehouse Drive, San Diego, 
CA 92121-1714, USA (e-mail: yunghsua@qti.qualcomm.com).}
\thanks{Haoran Hong is with University of Southern California, 3740 McClintock Ave.,
Los Angeles, CA 90089-2564, USA (e-mail: haoranho@usc.edu).}
\thanks{Gene Cheung is with York University, 4700 Keele Street, Toronto, Canada M3J 1P3 (e-mail: genec@yorku.ca).}
\thanks{Antonio Ortega is with University of Southern California, 3740 McClintock Ave.,
Los Angeles, CA 90089-2564, USA (e-mail: antonio.ortega@sipi.usc.edu).}
}

\maketitle
\begin{abstract}
An unfocused plenoptic light field (LF) camera places an array of microlenses in front of an image sensor in order to separately capture different directional rays arriving at an image pixel. 
Using a conventional Bayer pattern, data captured at each pixel is a single color component (R, G or B).
The sensed data then undergoes demosaicking (interpolation of RGB components per pixel) and conversion to an array of sub-aperture images (SAIs). 
In this paper, we propose a new LF image coding scheme based on graph lifting transform (GLT), where the acquired sensor data are coded in the original captured form without pre-processing. 
Specifically, we directly map raw sensed color data to the SAIs, resulting in sparsely distributed color pixels on 2D grids, and perform demosaicking at the receiver after decoding. 
To exploit spatial correlation among the sparse pixels, we propose a novel intra-prediction scheme, where the prediction kernel is determined according to the local gradient estimated from already coded neighboring pixel blocks. 
We then connect the pixels by forming a graph, modeling the prediction residuals statistically as a Gaussian Markov Random Field (GMRF). 
The optimal edge weights are computed via a graph learning method using a set of training SAIs. 
The residual data is encoded via low-complexity GLT. 
Experiments show that at high PSNRs---important for archiving and instant storage scenarios--- our method outperformed significantly a conventional light field image coding scheme with demosaicking followed by High Efficiency Video Coding (HEVC).
\end{abstract}

%
\IEEEpeerreviewmaketitle

\section{Introduction}
\begin{figure}[b]
\begin{center}
\includegraphics[width = .9\linewidth]{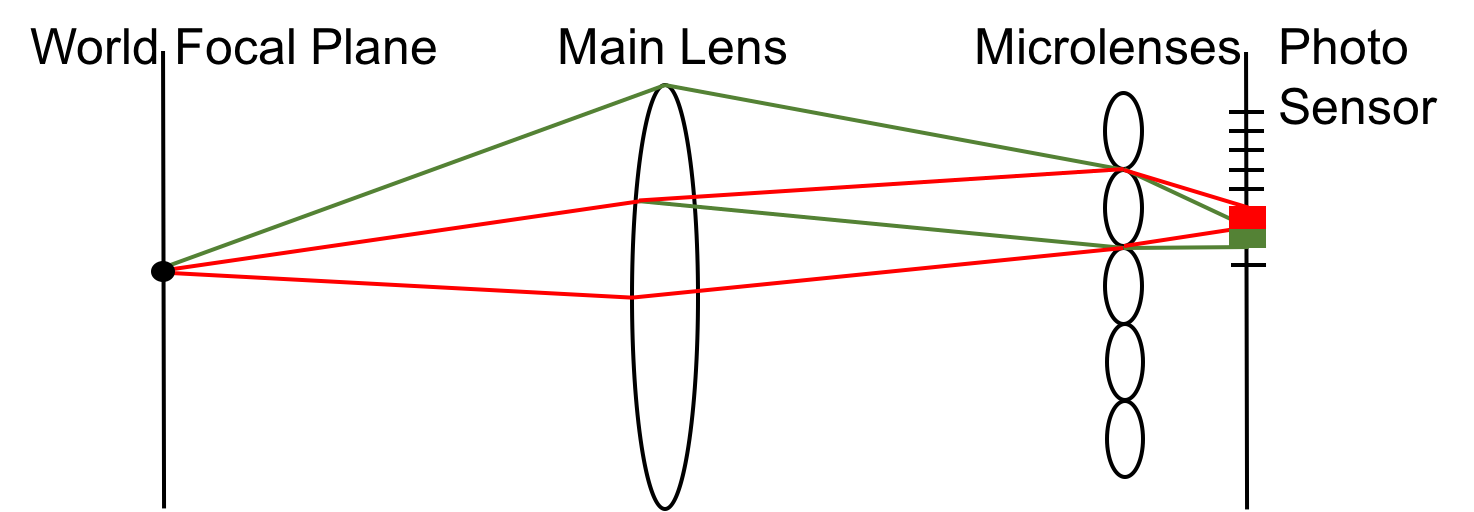}
\end{center}
\caption{Conceptual system of lenselet-based unfocused plenoptic 1.0 camera}
\label{fig:plenoptic-camera}
\vspace{-5mm}
\end{figure}
Light Field (LF) imaging separately captures light rays arriving from different directions at each pixel in an image. With acquired LF data, multi-view rendering \cite{levoy1996light,magnor2000data} and re-focusing \cite{ng2005light,levoy2004synthetic} become  possible post-capture.
However, captured LF data are very large in volume compared to a conventional color image of the same resolution, and hence efficient compression of LF data is important for storage and transmission. 
\begin{figure}[t]
\centering
\begin{subfigure}{.6\linewidth}
  \centering
  \includegraphics[width=\linewidth]{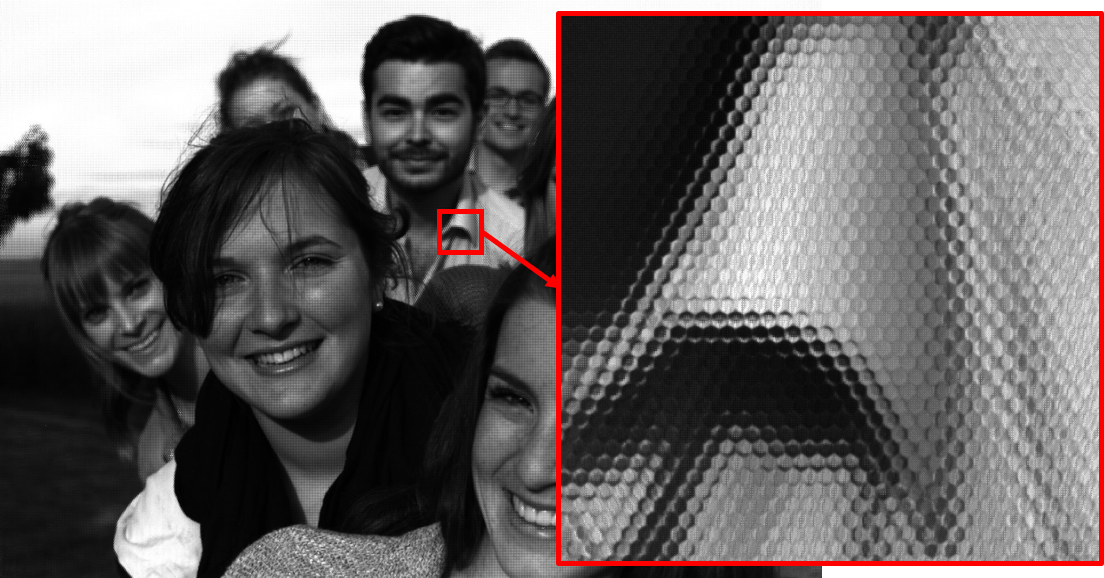}
  \caption{}
  \label{fig:sub_lenselet}
\end{subfigure}%
\hspace{1mm}
\begin{subfigure}{.6\linewidth}
  \centering
  \includegraphics[width=\linewidth]{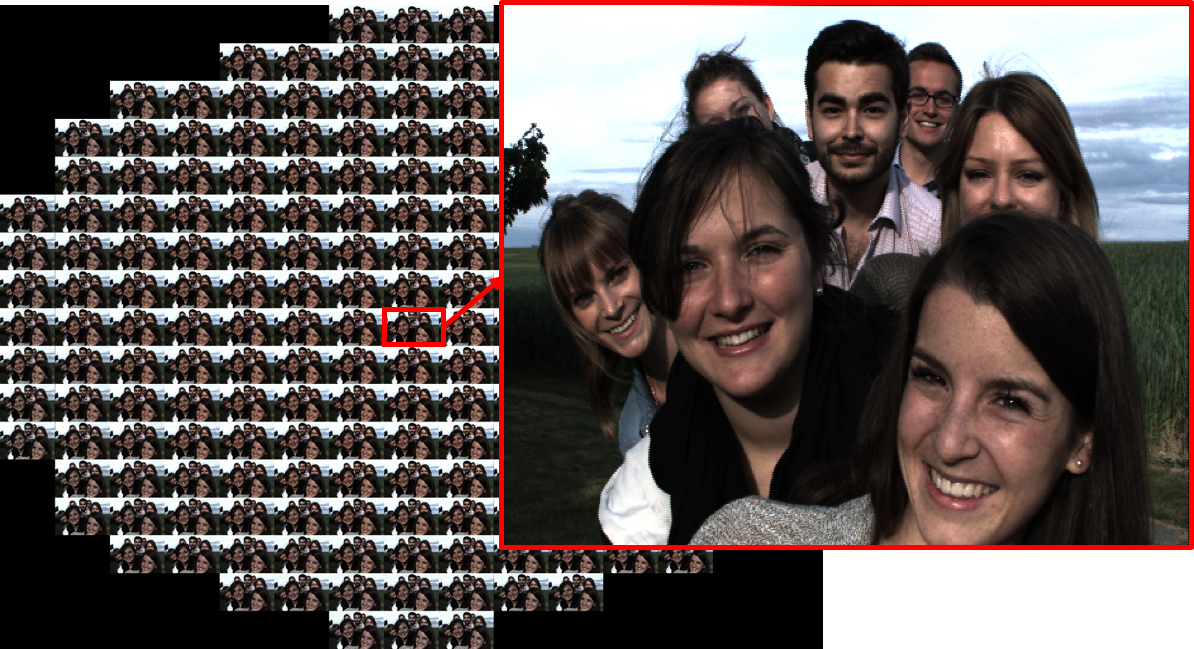}
  \caption{}
  \label{fig:sub_view}
\end{subfigure}
\vspace{-5mm}
\caption{\footnotesize (a) A raw lenselet image, \textit{Friends1}, from EPFL dataset and (b) the array of SAIs}
\label{fig:test}
\vspace{-5mm}
\end{figure}

In the last decade, many hardware designs have been developed for LF acquisition, including multiple camera arrays, aperture cameras, and lenselet-based plenoptic cameras. Among them, lenselet-based plenoptic cameras are the most popular. There are two types of plenoptic cameras that have been made commercially available: unfocused (or plenoptic 1.0), such as Lytro Illum \cite{Lytro}, and focused (or plenoptic 2.0), such as Raytrix R42 \cite{Raytrix}.

In an unfocused plenoptic camera, a microlens array is placed at a distance equal to the focal length in front of a conventional photo sensor using Bayer color filters, as shown in Fig. \ref{fig:plenoptic-camera}. 
The array produces a special hexagonal pattern on the resulting raw image (\textit{lenselet image}), as shown in Fig.\;\ref{fig:sub_lenselet}, where pixels within a hexagon correspond to light rays through a microlens. The obtained raw image is a 2D image comprising a 2D hexagonal grid of micro-images (MI). Since light fields are commonly represented and processed as 4D functions (\textit{i.e.}, 2D arrays of SAIs) \cite{gortler1996lumigraph,levoy1996light,ng2005light}, 
the acquired lenselet image typically undergoes 
{\em pre-processing}, consisting of demosaicking (pixelwise RGB interpolation) and conversion to multiple SAIs, as shown in Fig.\;\ref{fig:sub_view}. 
Each SAI is like a typical 2D image, gathering pixels from a specific light direction.

There exist two types of redundancies within the 4D LF representation: i) spatial redundancy among neighboring pixels in an SAI, i.e., \textit{intra-view correlation}, and ii) angular redundancy among SAIs in a neighborhood, i.e., \textit{inter-view correlation}. 


In \cite{conti2012new, conti2016hevc, monteiro2016light}, intra- and inter-view correlations are exploited using regular HEVC intra-prediction modes, together with a new \textit{self similarity} (SS) mode, which is similar to Intra Block Copy (IBC) in screen content coding extension of HEVC (HEVC-SCC).
In \cite{Dubreuil2019,Viola2018,WaSP}, view synthesis methods are used to recover the entire light field from a small subset of encoded views. 
In \cite{dai2015lenselet,liu2016pseudo, zhao2016light, hariharan2017low,vieira2015data, li2017pseudo,Jia2017,MONTEIRO2021}, Pseudo Video Sequence (PVS)-based coding schemes are proposed, where the SAIs are treated as frames in a video  sequence, so that intra- and inter-prediction tools in the video codec can be used.
In \cite{Wang2016, Mehajabin2019, Khoury2019}, inter-view prediction using multiview extension of HEVC (MV-HEVC) is proposed to address the loss of correlation among views caused by scanning in PVS-based schemes. In \cite{Ricardo2020}, a hybrid data representation method is proposed to simultaneously exploit intra- and inter-view correlations in SAIs and MIs.
Other state-of-the-art methods can be found in recent surveys \cite{Survey2020, Survey2021}.



While exploiting inter-view correlations results in higher coding efficiency, it also leads to more complex encoding (due to motion / disparity prediction) and creates dependencies among coded SAIs, which is undesirable for random access. 
In particular, in an archiving scenario, a user may desire to quickly browse through viewpoint images, each of which can be synthesized in acceptably high quality using only a small subset of SAIs \cite{piexoto2017}. 
Thus, speedy extraction of this image subset from the LF data compressed in high quality is important. 
Moreover, synthesizing viewpoint images with a small set of SAIs also reduces memory requirement, as the number of images needed to be decoded and buffered is reduced.
Finally, we note that, for initial storage, most standard digital cameras use a low complexity codec (JPEG) operating by default at very high PSNR (see Section \ref{sec:2nd-analysis}). 
In analogy, in this paper, we consider an \textit{intra-view-only} approach (which leads to faster encoding and better random access), operating at high rates / PSNR. 

Another source of inefficiency in existing techniques is the redundancy generated by color interpolation in the demosaicking process, which increases the number of pixels to be encoded. Moreover, in methods utilizing existing image / video codecs such as JPEG, H.264 and HEVC, input RGB image/videos are typically converted to other representations with chroma downsampling, \textit{e.g.} 4:2:0 YUV format, which introduces distortions due to integer rounding and color sub-sampling. 

To avoid the demosaicking process that generates redundancy before compression, in our preliminary work \cite{chao2017pre}, we proposed a new pre-processing pipeline, where the original lenselet color pixels captured by the photo sensor are mapped onto sparse locations in a series of SAIs without performing demosaicking and color conversion prior to encoding. 
Unlike pre-demosaic compression proposed for regular image and video in \cite{lee2001novel,koh2003new,chung2008lossless,lee2009novel}, where RGB pixel locations follow the regular Bayer filtering pattern, pixels in  pre-demosaic SAIs are irregularly placed, making it difficult to apply existing standard image and video coding schemes, \textit{e.g.,} JPEG and HEVC. 
To resolve this difficulty, we proposed placing pixels belonging to each SAI on a graph and encoding them using a graph-based lifting transform \cite{narang2009lifting}. 
Since a pixel only connects to other pixels in the same SAI, pre-demosaicked pixels in different SAIs can be encoded and decoded independently, which favors random access efficiency. 

In this paper, we develop a complete system based on the preliminary ideas presented in \cite{chao2017pre}. 
Specifically, we introduce a novel intra-prediction approach that can operate on the pre-demosaic image, achieving significant improvements in coding efficiency. Additionally, we develop a systematic approach to optimize edge weights for the graph to be used for our graph-based lifting transform. 
 


For intra-prediction, we propose a novel intra-prediction strategy designed for sparsely distributed pixels in SAIs. 
Specifically, we first estimate the local characteristics, \textit{e.g.}, object contour, of each block based on \textit{structure tensor} \cite{Bigun87optimalorientation}, using already encoded neighboring pixel blocks. 
The information is then used to adjust the shape of kernels used for prediction. 
As a result, 
the predictor will use more information from neighboring pixels on the same side of a contour and less information from pixels across the contour, preserving contour sharpness after prediction.


The graph weights, which reflect similarities between two connected pixels of residual intensities, are optimized based on Gaussian Markov Random Field (GMRF) modeling of the signal. 
The problem of learning efficient graph structures, \textit{i.e.}, graph topology and weights, from training data in image and video has been studied recently  \cite{hu2015intra, pavez2015gtt, pavez2016generalized, egilmez2016gbst}. However, most methods are designed for regularly sampled data, \textit{e.g.}, pixels on a rectangular grid and therefore, unsuitable for irregularly distributed pixels. In this work, we consider a graph learning problem where graphs is derived from training vectors with missing entries that are different from vector to vector. 

After defining a graph, the graph signals, \textit{i.e.}, the residual pixel intensities, are encoded using a graph-based lifting transform \cite{narang2009lifting}. 
Compared to the state-of-the-art HEVC-based coding, the results show significant gain---around $5dB$ on average---at the high bitrate/PSNR range, which is important for applications that demand high-quality encoding such as archives. 
Compared to our preliminary work \cite{chao2017pre}, we observe around $3dB$ gain at both high and low bitrates. 


The outline of the paper is as follows. 
In Section \ref{background}, we present notations used throughout this paper and review conventional approaches in lenselet-based LF image compression. 
We provide an overview of our proposed coding scheme in Section \ref{proposed-prepip}. 
In Section \ref{prediction} and \ref{transform-coding}, we describe in details our proposed intra-prediction and transform coding algorithms tailored for sparsely distributed pixels. Experiments and conclusions are presented in Section \ref{Experiments} and \ref{conclusion}, respectively.

\section{Notations and Background}
\label{background}
\subsection{Notations}
\begin{table}
\centering
\captionof{table}{List of notations ant their meanings}
\resizebox{\linewidth}{!}{%
  \begin{tabular}{ | c || c |}
    \hline
    \textbf{Notations} & \textbf{Meanings} \\ \hline \hline
    $f(\cdot), K(\cdot)$  & functions  \\ \hline
    $f(i),  K(i)$ & function value at location $i$ \\ \hline
    $\nabla \fv $ & gradient vector of function $f(\cdot)$ \\ \hline
    $f_i$ & $i_{th}$ element in vector $\fv$ \\ \hline
    $F_{i,j}$ & element of matrix $\Fm$ at position $(i,j)$  \\ \hline
    $\fv_\mathcal{S}$ & subvector of $\fv$ extracted from positions specified by set $\mathcal{S}$ \\ \hline
    $\Fm_{\mathcal{S},\mathcal{P}}$ & submatrix of $\Fm$ extracted from positions specified by set $\mathcal{S}$ and $\mathcal{P}$ \\ \hline
    $\Fm^{\transp}$ & matrix transpose \\ \hline
    $\Fm^{-1}$ $|$ $\Fm^{\dagger}$ & inverse $|$ pseudo-inverse of $\Fm$ \\ \hline
    $x \sim \mathcal{N}(\muv, \Sigmam)$ & multivariate Gaussian with mean $\muv$ and covariance $\Sigmam$ \\ \hline 
    $\Qm$ & inverse covariance (precision) matrix of multivariate Gaussian \\ \hline    
    $O(\cdot)$ & big O notation in complexity analysis \\ \hline
    $\Am$ $|$ $\Lm$ & graph adjacency $|$ Laplacian matrix\\ \hline
    $\Dm$ $|$ $\Hm$ & graph degree $|$ self loop matrix \\ \hline   
  \end{tabular}  }
  \label{tab:notation}
\end{table}
Throughout this paper, we use lowercase normal (\textit{e.g.,} $f$ and $\sigma$), lowercase bold (\textit{e.g.,} $\fv$ and $\sigmav$) and uppercase bold letter (\textit{e.g.,} $\Fm$ and $\Thetam$) to denote scalars, vectors, and matrices, respectively.
The elements of a $2$-dimensional vector (\textit{e.g.,} $\vv = [v_1, v_2]^{\transp}$) in the Euclidean space represent the directions along the horizontal and vertical axis respectively, a notation commonly used for the image coordinate system. 
Unless otherwise stated, calligraphic capital letters (e.g., $\mathcal{S}$) are used to represent sets. 
Other notations are summarized in Table \ref{tab:notation}.
\begin{figure*}[t]
\begin{center}
\includegraphics[width = 0.8\textwidth]{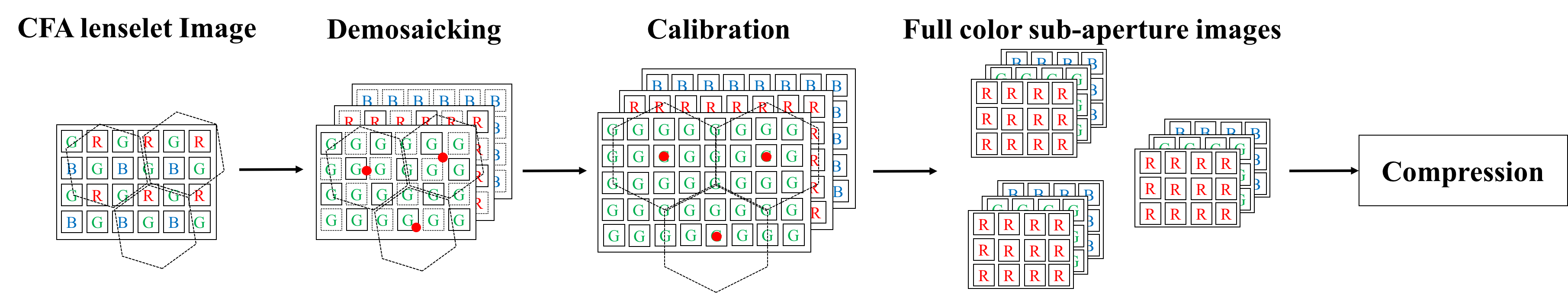}
\end{center}
\caption{A conventional pre-processing pipeline for light field image compression is shown. The demosaicking and calibration processes are applied before compression. Macro-pixels are represented by dashed lines, and macro-pixel centers are denoted by red points.}
\label{fig:traditional-pip}
\end{figure*}
\subsection{Conventional pre-processing pipeline before compression}
Fig.\;\ref{fig:traditional-pip} shows a block diagram of pre-processing steps before a conventional coding scheme is applied on light field image, where the captured Bayer-patterned lenselet image is converted into a 2D array of full-color SAIs. 
In the figure, the conversion process is based on the method proposed by Dansereau et al. \cite{dansereau2013decoding, LFtoolbox}. 

First, in order to generate full-color lenselet images, the missing color components at each pixel are interpolated using nearby pixels containing the target colors. 
The number of pixels will be increased threefold through this process regardless of the demosaicking algorithm used.

Second, the color lenselet image is calibrated via rotation, translation and scaling, which requires interpolation and increases the data size.
A lenselet image consists of multiple hexagonally arranged pixel patches called \textit{macro-pixels}. Each macro-pixel collects light for one image pixel arriving from different directions. 
However, due to manufacturing defects, the arrangement of macro-pixels is usually not aligned with the image coordinates.
So, the color lenselet image needs to be calibrated with the help of white images \cite{dansereau2013decoding}, so that each macro-pixel center falls onto an integer pixel location, and the arrangement of macro-pixels is aligned to the regular hexagonal grid. 

Third, pixels that correspond to light rays coming from the same direction are arranged into one SAI. 
After the arrangement, the pixel grid is corrected to be square. The resampling step, which involves interpolation \cite{dansereau2013decoding}, increases the data size again. 

\section{Proposed pre-processing pipeline for graph based compression}
\label{proposed-prepip}
\begin{figure*}[t]
\begin{center}
\includegraphics[width = .7\textwidth]{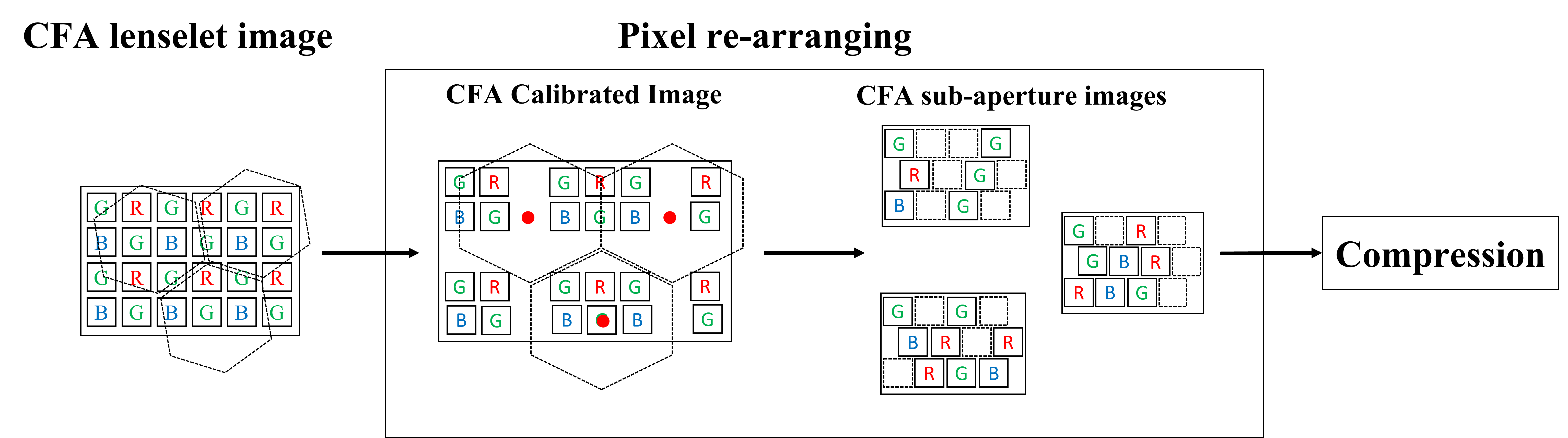}
\end{center}
\caption{Proposed pre-processing pipeli{\color{red}n}e for the color filter array (CFA) of lenselet image}
\label{fig:proposed}
\end{figure*}
\begin{figure}[htbp]
\begin{center}
\includegraphics[width = 0.7\linewidth]{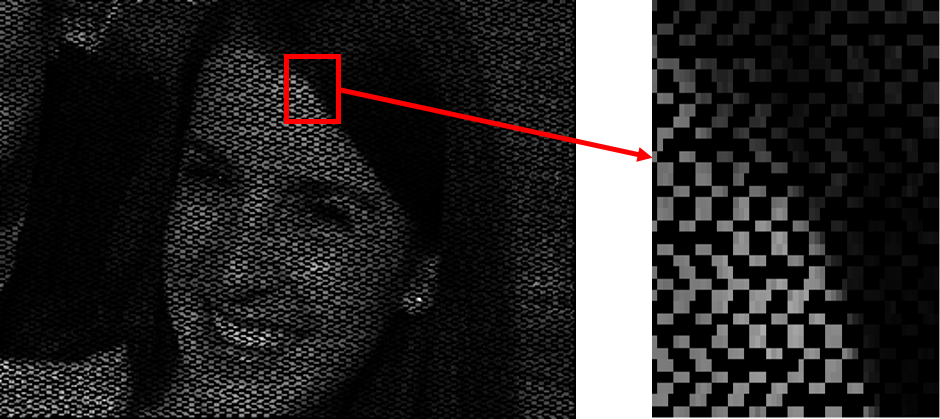}
\end{center}
\caption{Since no interpolation, i.e., color interpolation in demosaicking and scaling in calibration, is applied, some pixel locations are empty in the SAIs.}
\label{fig:crop_subaperture}
\end{figure}
To avoid the aforementioned redundancies, we propose a new pipeline to pre-process LF data, shown in Fig.\;\ref{fig:proposed}.

First, we postpone demosaicking and calibrate raw lenselet images via affine transform to align all lenselet centers with pixel centers.
Pixels that fall onto non-integer locations after transformation will be moved to the nearest integer positions.

Then, based on the relative locations within the macro-pixels on the calibrated image, pixels are arranged onto multiple SAIs. 
Due to the placement of macro-pixels, the pixels in the SAIs are placed hexagonally, which is different from the conventional pipeline, where SAIs are re-sampled into rectangular placement. 

Note that the pixels around the boundaries of each macro-pixel, which tend to have noisy values due to underexposure, are discarded in the pipeline described in \cite{dansereau2013decoding}. 
However, in the proposed pipeline, the boundary pixels are kept. 
Our pipeline, which does not change the number of pixels nor the RGB values in the raw data, is reversible. Thus, if lossless compression is applied the original lenselet images can be recovered. 

An example of sparsely distributed G components on one SAI is shown in Fig.\;\ref{fig:crop_subaperture}. 

The pattern of available pixels on each SAI depends on the inputed white image that is selected based on the camera
zoom and focus settings appropriate for the lenselet picture to be decoded.
Hence, the pattern may change as camera settings change \cite{LFtoolbox}.
Additionally, the pattern may also change for different types of macro-pixel misalignments and different calibration algorithms adopted. 

We stress that our scheme described in the next few sections does not rely on a particular selection of algorithms for demosaicking and calibration. 
In fact, many different techniques have been proposed in recent years for LF image calibration \cite{xu2014multi,bok2017geometric} and demosaicking \cite{dansereau2013decoding,xu2014multi,seifi2014disparity}. 
Our coding scheme can be easily adapted to these various approaches by using different calibration transforms at the encoder for pixel rearrangement. The demosaicking strategy at the decoder side can also be adjusted accordingly.

\section{Proposed Intra-prediction scheme}
\label{prediction}

\begin{figure}[t]
\begin{center}
\includegraphics[width = 0.9\linewidth]{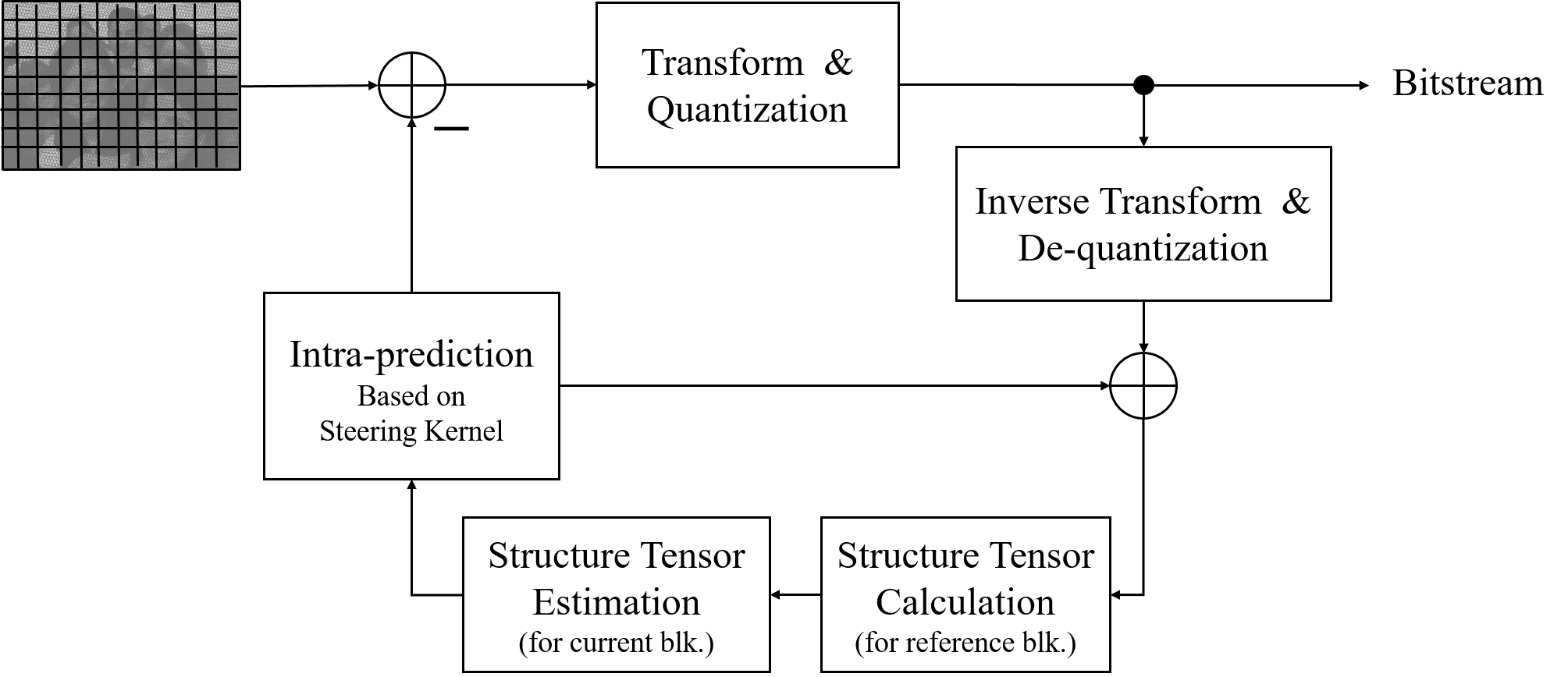}
\end{center}
\caption{Proposed Intra-prediction system for sparsely distributed pixels}
\label{fig:intra-system}
\vspace{-5mm}
\end{figure}

After calibration and decomposition, each SAI containing sparsely distributed pixels undergoes our proposed intra-prediction. 
Fig.\;\ref{fig:intra-system} shows a flow chart of our intra-prediction scheme.
First, each SAI is divided into non-overlapped $m\times m$ blocks for intra-prediction, which is the basic unit for later transform coding.
Second, we estimate gradients of all the pixels in each reference block, using which we compute the structure tensor for the block. 
Third, structure tensors of decoded neighboring blocks are used to estimate gradient direction and strength of a target block to be encoded, from which we can determine the parameters of an \textit{adaptive kernel} \cite{takeda2007kernel} used for intra-prediction. 
Hence, no overhead is required in our intra-prediction scheme, as the choice of a prediction mode is solely based on information from previously encoded blocks. 

\subsection{Gradient Estimation}
\label{sec:gradient-calc}

In our work, the local gradient at each sparse pixel $i$, $\nabla \fv(i)$, is estimated via \textit{linear regression}. 
Using Taylor series expansion, the pixel value at a given point $\xv$ can be expressed as 
\begin{equation}
f(\xv) = f(\av) + \nabla \fv(\av)\cdot (\xv-\av) + \mathcal{O}(\|\xv-\av\|^2)
\end{equation}
where $\nabla \fv(\av)$ is the gradient vector at point $\av$. For $\av$ sufficiently close to $\xv$, the pixel value $f(\xv)$ can be well approximated using only the first two terms 
\begin{equation}
f(\xv) \approx f(\av) + \nabla \fv(\av)\cdot (\xv-\av)
\label{equ:linear-approx}
\end{equation}
Based on the linear approximation, we estimate the gradient $\nabla \fv(\av)$ at point $\av$ by fitting a hyperplane which best satisfies Equ.\;\eqref{equ:linear-approx} for a number of pixels $\{\xv_1, \xv_2, \cdots \xv_k\}$ close to $\av$. The fitting can be represented as an overdetermined system:
\begin{align}
\textbf{F} &= \textbf{X}\cdot \nabla \fv(\av) \\
&= \begin{bmatrix}
f(\xv_1) - f(\av) \\
f(\xv_2) - f(\av) \\
\vdots \\
f(\xv_k) - f(\av)
\end{bmatrix}
= \begin{bmatrix}
(\xv_1 - \av)^{\transp} \\
(\xv_2 - \av)^{\transp} \\
\vdots \\
(\xv_k - \av)^{\transp}
\end{bmatrix}\cdot \nabla \fv(\av) 
\label{equ:overdetermined}
\end{align}
The optimal gradient $\nabla \fv(\av)^*$ can be computed by solving a least-square problem 
\begin{equation}
\nabla \fv(\av)^* = \argmin_{\nabla \fv(\av)} \| \textbf{F} - \textbf{X}\cdot \nabla \fv(\av)\|_2^2 
\end{equation}
which has a closed-form solution
\begin{equation}
\nabla \fv(\av)^* = (\textbf{X}^{\transp}\textbf{X})^{\dagger}\textbf{X}^{\transp}\textbf{F}.
\end{equation}
For each pixel, we use four nearby pixels ($k=4$) for hyperplane fitting in Equ.\;\eqref{equ:overdetermined}, namely, the two closest neighbors in horizontal and vertical orientations respectively. 
Selecting close neighbors along different directions alleviates the problem when all pixels used for fitting are aligned, resulting in an ill-conditioned system.

With the obtained gradient of available pixels, the structure tensor of block $\text{B}$ is calculated as: 
\begin{equation}
\Hm_\text{B} = \Sigma_{i\in \text{B}}\nabla \fv(i)\nabla \fv(i)^{\transp}
= \Sigma_{i\in \text{B}}\begin{bmatrix}
  d_h(i)^2 & d_h(i)d_v(i) \\
  d_v(i)d_h(i) & d_v(i)^2
 \end{bmatrix} 
 \label{equ:structure-tensor}
\end{equation}
where $d_h(i)$ and $d_v(i)$ denote the gradients of pixel intensities at pixel $i$ along horizontal and vertical directions, respectively. 

After computing the structure tensor in Equ.\;\eqref{equ:structure-tensor} using the estimated gradients, we apply eigen-decomposition on the $2\times 2$ matrix $\Hm_\text{B}$:
\begin{align}
\Hm_\text{B} &= \begin{bmatrix}
\ev_1 & \ev_2
\end{bmatrix}\begin{bmatrix}
\lambda_1 & 0 \\
0 & \lambda_2
\end{bmatrix}\begin{bmatrix}
{\ev_1}^{\transp} \\
{\ev_2}^{\transp}
\end{bmatrix} \\
&= \lambda_1 \ev_1 {\ev_1}^{\transp} + \lambda_2 \ev_2 {\ev_2}^{\transp}
\label{equ:tensor-eigen-decomp}
\end{align}
The corresponding eigenvectors $\ev_1$ and $\ev_2$ along with their eigenvalues $\lambda_1$ and $\lambda_2$,  where $\lambda_1 < \lambda_2$, summarize the gradient distribution within block $\text{B}$. 
Eigenvector $\ev_2$ represents the direction maximally aligned with the block gradient, while the orthogonal $\ev_1$ represents the dominant edge direction. 
The two eigenvalues $\lambda_1$ and $\lambda_2$ provide information of gradient strength along directions $\ev_1$ and $\ev_2$, which will be used in constructing adaptive kernels for intra-prediction (Section \ref{sec:adaptive-kernel}).  

\subsection{Structure Tensor Estimation}

\begin{figure}[t]
\begin{center}
\includegraphics[width = 0.9\linewidth]{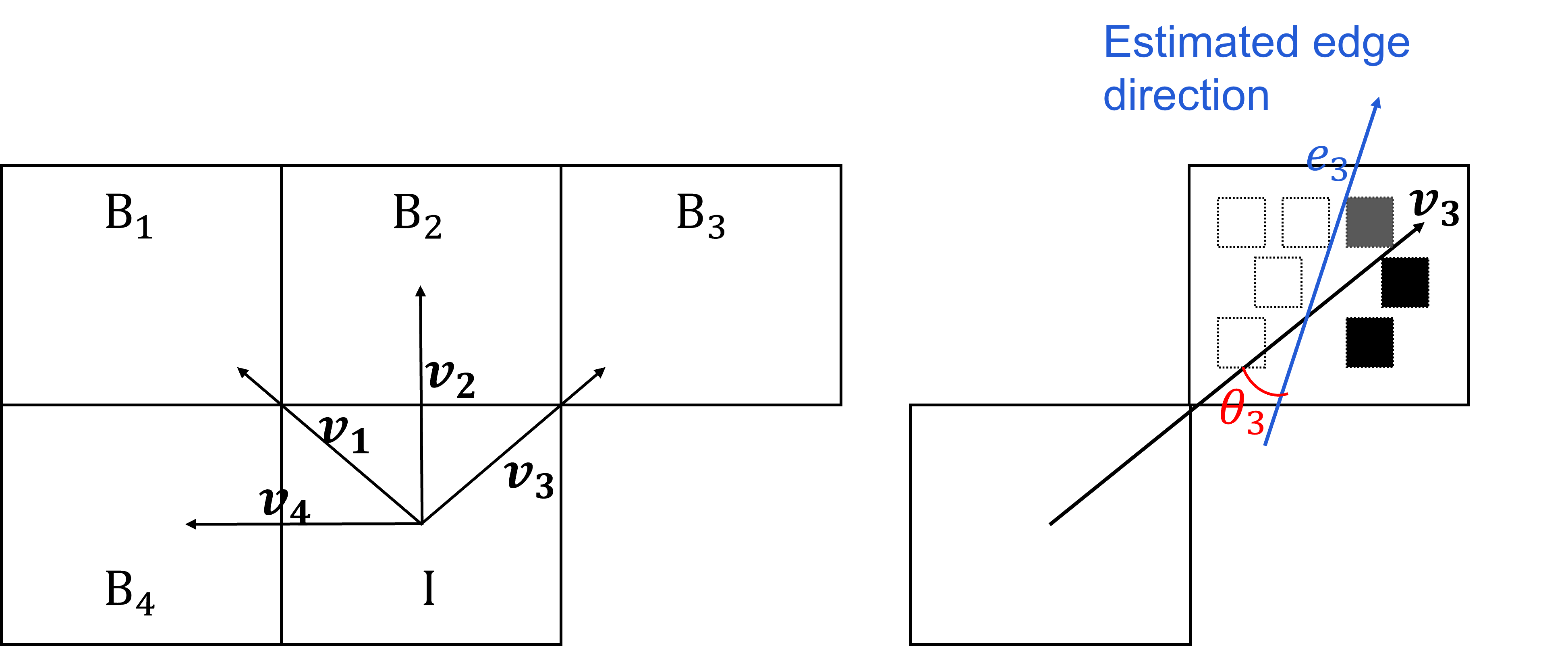}
\end{center}
\caption{4 decoded reference blocks and the vectors indication their relative locations to the input block $\text{I}$ and edge direction estimated}
\label{fig:intra-ref}
\end{figure}

The estimate of structure tensor $\Hm_\text{I}$ of input block $\text{I}$ is calculated as the weighted average of the structure tensors from its decoded neighboring blocks $\{\text{B}_1, \text{B}_2, \cdots \}$ (See Fig.\;\ref{fig:intra-ref}). 
The estimate is written as
\begin{equation}
\Hm_\text{I} = \dfrac{1}{c} \Sigma_i w_i \cdot (\dfrac{1}{n_i} \Hm_{\text{B}_i}),
\label{equ:structure-estimation}
\end{equation}
where $n_i$ denotes the number of available pixels in reference block $\text{B}_i$, $w_i$ is the weight associated with block $\text{B}_i$, and $c$ is the normalization constant $c = \Sigma_i w_i$.

Each weight value $w_i$ is a function of the cosine similarity between the gradient direction in reference block $\text{B}_i$ and the relative direction from the input block $\text{I}$ to $\text{B}_i$. 
The design is based on the observation that edges in natural images are continuous contours, and can be approximated locally with straight lines. 
In other words, the orientation of edges is mostly consistent locally. 
Hence, if the edge direction calculated on one reference block $\text{B}_i$ is consistent with its relative location from $\text{I}$, denoted by a vector $\vv_i$ from $\text{I}$ to $\text{B}_i$ as shown in Fig.\;\ref{fig:intra-ref}, then block \text{I} is more likely to have the same edge orientation. 
In estimating the gradient in the input block, reference blocks with consistent edge orientation are assigned larger weights.

In our algorithm, we consider the four decoded neighboring blocks shown in Fig.\;\ref{fig:intra-ref} as references for structure tensor estimation for intra-prediction.  
Denote by $\vv_i$ the (unit norm) vector from $\text{I}$ to reference block $\text{B}_i$, $\ev_i$ the edge direction computed from structure tensor in block $\text{B}_i$, and $\theta_i$  the angle between $\vv_i$ and $\ev_i$ (Fig.\;\ref{fig:intra-ref}), calculated as
\begin{equation}
\theta_i = \arccos\left(\frac{\vv_i\cdot \ev_i}{\|\vv_i\|_2}\right)  \in [0, \pi],
\end{equation}
with 
\[
\{\vv_1, \vv_2, \vv_3, \vv_4\} = \{[\frac{-1}{\sqrt{2}},\frac{-1}{\sqrt{2}}]^{\transp}, [-1, 0]^{\transp}, [\frac{-1}{\sqrt{2}},\frac{1}{\sqrt{2}}]^{\transp}, [0,-1]^{\transp}\}.
\]
The weight value $w_i$ in Equ.\;\eqref{equ:structure-estimation} is defined as 
\begin{equation}
w_i =
\begin{cases}
    \exp(-\frac{\theta_i}{\delta}) &  \text{if } \theta < \frac{\pi}{2} \\
    \exp(-\frac{\pi - \theta_i}{\delta})  &  \text{if } \theta > \frac{\pi}{2}\\
\end{cases}
\label{equ:tenstor-estimation-weights}
\end{equation}
where $\delta$ is a parameter adjustable by users. Larger weights are assigned to blocks with smaller $\theta_i$, \textit{i.e.} such that the edge direction $\ev_i$ and unit vector $\vv_i$ are nearly consistent.

Note that the pixel intensities $\fv$ considered in the gradient calculation in Section \ref{sec:gradient-calc} and the structure tensor estimation described in Equ.\;\eqref{equ:structure-estimation} are based on only G components, since the sampling rate of G components is higher than for B and R components in the Bayer pattern.
This leads to a more accurate approximation in Equ.\;\eqref{equ:linear-approx}. 
The structure tensor and  kernel parameters obtained from G pixels (see next section) are applied to B and R components given that 
there is a high correlation among the intensity variations in different color channels. 
\begin{figure}[t]
\begin{center}
\includegraphics[width = 0.85 \linewidth]{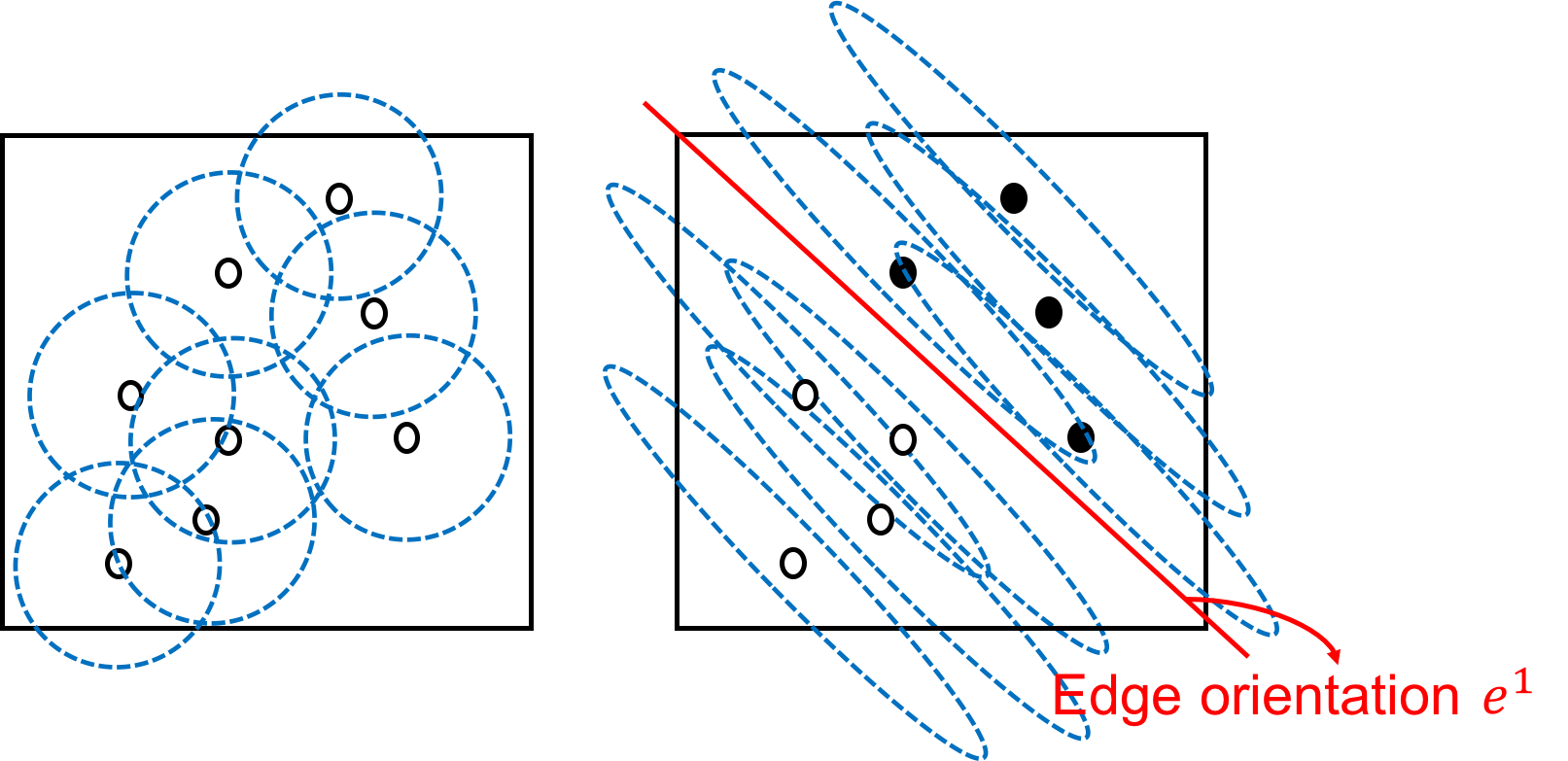}
\end{center}
\caption{Illustration of kernel size and shape in the smooth block (left) and the block with strong edge (right)}
\label{fig:kernel}
\end{figure}

\subsection{Data-adaptive Kernel Regression}
\label{sec:adaptive-kernel}

For intra-prediction, we apply an adaptive kernel regression \cite{takeda2007kernel} centered at each pixel. 
In the case of zero-order estimation, a prediction of the pixel intensity at $\xv$ can be calculated as the weighted average of its neighboring pixels $\xv_i$, written as
\begin{equation}
\tilde{f}(\xv) = \frac{1}{z_{\xv}}\Sigma_{\xv_i\in \mathcal{R}_\xv} (K(\xv_i-\xv)\cdot f(\xv_i))  
\end{equation}
where $\mathcal{R}_\xv$ denotes the set of neighboring pixels of $\xv$ in the decoded reference blocks, and $z_{\xv}$ is the normalization constant $z_{\xv} = \Sigma_{\xv_i\in \mathcal{R}_\xv} K(\xv_i-\xv)$. A common choice for the kernel function $K(\cdot)$ is the Gaussian kernel.
For the  data-adaptive kernel regression used in this work, the Gaussian kernel is adapted based on the edge orientation and strength derived from the computed structure tensor as
\begin{equation}
K({\xv_i-\xv}) = \frac{\sqrt{\det(\bf C)}}{2\pi\sigma^2}\exp(-\frac{(\xv_i-\xv)^{\transp} \Cm (\xv_i-\xv)}{2\sigma^2}),
\label{equ:adaptive-Gaussian-kernal}
\end{equation}
where ${\bf C}$ is the covariance matrix based on the eigenvectors and eigenvalues of structure tensor from Equ.\;\eqref{equ:tensor-eigen-decomp}, and can be decomposed as
\begin{align}
\textbf{C} &= \phi\cdot \begin{bmatrix}
\ev_1 & \ev_2
\end{bmatrix}\Lambda \begin{bmatrix}
{\ev_1}^{\transp} \\ 
{\ev_2}^{\transp}
\end{bmatrix} \\
\label{equ:weight-numerical-stable}
\Lambda &= \begin{bmatrix}
\frac{1}{\epsilon} & 0 \\
0 & \epsilon
\end{bmatrix} \quad \text{, }\epsilon = \frac{\lambda_2 + p_2}{\lambda_1 + p_2}\\
\phi &= \frac{1}{n}\sqrt{\lambda_1\lambda_2 + p_1},
\end{align}
and where $[\ev_1 \quad \ev_2]$ rotates the coordinates of Gaussian kernel along the dominant gradient and edge directions, and $\epsilon$ is the ratio of two eigenvalues, representing the relative strength of gradient in $\ev_2$ from the perpendicular direction $\ev_1$. The kernel will be elongated for blocks of strong edges, where $\lambda_2 \gg \lambda_1$, and will be near-circular for smooth block ($\lambda_2 \approx \lambda_1 \approx 0$), as illustrated in Fig.\;\ref{fig:kernel}. $\phi$ determines the scaling of the kernel size, where $n$ is the number of available pixels in the block and $p_1$ and $p_2$ are two positive scalars used to ensure numerical stability.


\section{Proposed transform coding} 
\label{transform-coding}
After intra-prediction, 
the residuals are placed on the vertices of the graphs optimized by the training set and then undergo graph-based lifting transform. 

\subsection{Preliminaries for Graphs}

A graph $\mathcal{G=(V,E)}$ contains a set of $n$ \textit{vertices} $\mathcal{V}=\{(v_i, h_i)\,|\, 1 \leq i \leq n\}$, and a set of $m$ \textit{links} $\mathcal{E} = \{(e_{i,j}, w_{i,j})\,|\, 1 \leq i,j \leq n\}$, each connecting two distinct nodes $v_i$ and $v_j$. The value $h_i \geq 0$ denotes the vertex weight, also called \textit{self loop}. For a \textit{weighted graph}, $w_{i,j}$ is a non-negative real-valued weight on $e_{i,j}$, which captures the similarity between the two connected nodes. 
We can represent the graph connectivity information using a $n\times n$ \textit{adjacency matrix} $\Am$, which has zero diagonal elements and $\Am_{i,j} = w_{i,j}$ for off-diagonal terms. A \textit{degree matrix} $\Dm$ is a $n\times n$ diagonal matrix with $D_{i,i} = \sum_j A_{i,j}$. A \textit{self loop matrix} $\Hm$ is a diagonal matrix with elements $H_{i,i} = h_i$. A \textit{generalized Laplacian matrix} is defined as $\Lm = \Dm-\Am+\Hm$. A \textit{combinatorial Laplacian matrix} is a special case of the general Laplacian matrix with no self loops. 
\subsection{Graph Learning for sparsely distributed pixels}
\label{sec:graph-learning}

In this section, we construct an \textit{optimal} graph that best describes inter-pixel similarities for each mode given observed intra-predicted residual blocks in LF images, assuming a known statistical model.

LF images are divided into a training set and a test set. Blocks of intra-predicted residuals from the training data are classified, based on their respective computed structure tensor, into $9$ modes, including $8$ directional modes:
 \begin{equation}
 \Mc = \{\Mc_\theta|\quad\theta = \frac{-3\pi}{8}, \frac{-\pi}{4}, \frac{-\pi}{8}, 0, \frac{\pi}{8}, \frac{\pi}{4}, \frac{3\pi}{8}, \frac{\pi}{2}\}
 \label{equ:training-direction}
 \end{equation}
 and a DC mode, $\Mc_\text{DC}$, if there is no dominant edge direction. 
Given the edge angle $\gamma$ of a training block (the angle between the smallest eigenvector and the horizontal axis) and the eigenvalues of its structure tensor ($\lambda_1$ and $\lambda_2$, where $\lambda_1\leq \lambda_2$), the class of the associated residual block is determined as
 \begin{equation}
 \Mc = \begin{cases}
     \Mc_\text{DC} &  \text{if } \frac{\lambda_2 + p}{\lambda_1 +p} < T \\
     \Mc_\theta  &  \text{if } \gamma \in [\theta-\frac{\pi}{16},  \theta+\frac{\pi}{16})
 \end{cases}.
 \label{equ:training-classification}
\end{equation} 
Training blocks in a class are assumed to be samples of the same statistical model, for which the corresponding \textit{optimal} graph is derived. The problem of finding the optimal graph can be seen as that of finding an optimal graph Laplacian matrix, which completely characterizes the graph. Hence a total of $9$ $n\times n$ 
generalized 
graph Laplacian matrices $\Lm_{i,i = 1, 2,\cdots 9}$ are derived through learning from intra-predicted residuals in the training set.


In the literature, multiple techniques have been proposed to find the optimal graph Laplacian matrix in different contexts. 
In \cite{pavez2015gtt,egilmez2016gbst,pavez2016generalized,dong2016learning,egilmez2016graphlearning}, a statistical assumption was made on data, where each observation $\fv\in \mathbb{R}^N$ is modeled as a realization of a \textit{Gaussian Markov Random Field} (GMRF), \textit{i.e.}, $\fv \sim \mathcal{N}(\muv, \Qm^{-1})$, with probability density function
\begin{equation}
p(\fv|\muv,\Qm) = \frac{\det(\Qm)^{\frac{1}{2}}}{(2\pi)^{\frac{N}{2}}}\exp\bigg(-\frac{1}{2}(\fv-\muv)^{\transp}\Qm(\fv-\muv)\bigg), 
\label{equ:GMRF-definition}
\end{equation}
where $\muv$ is the mean vector, and $\Qm$ is the inverse covariance matrix (precision matrix). 
The problem for learning the graph Laplacian matrix $\Lm$ given the GMRF assumption is to find the maximum likelihood estimate of $\Qm$, which defines the partial correlation between pairwise variables in $\fv$:
\begin{equation}
p(f_i,f_j\,|\,\fv \setminus \{f_i,f_j\}) = -\frac{Q_{i,j}}{\sqrt{Q_{i,i}Q_{j,j}}}
\end{equation}
In \cite{egilmez2016gbst,egilmez2016graphlearning}, Egilmez et al. proposed an algorithm specifically to compute a precision matrix $\Qm$ targeting a graph Laplacian structure with non-negative edge weights, \textit{i.e.},  
\begin{equation}
\begin{cases}
   Q_{i,j} < 0 \quad \text{if } A_{i,j} > 0 \\
   Q_{i,j} = 0 \quad \text{if } A_{i,j} = 0 \\
\end{cases}
\label{equ:laplacian_struct}
\end{equation}
with additional constraints on the graph connectivity.
Given $l$ \textit{i.i.d.} observations $\{\fv_1, \fv_2, \cdots \fv_l\}$ of zero mean GMRF, the precision matrix can be derived by solving the maximum likelihood (ML) problem:
\begin{equation}
\begin{aligned}
&\argmax_{\Qm} \prod_{i=1:l} p(\fv_i|\Qm,\muv_i=\textbf{0}) \\
= &\argmin_{w,v} \quad \log\det(\Qm) - \Tr(\Qm\Sm)  \\
&\text{subject to  } \Qm = \Bm\dg(\wv)\Bm^{\transp} + \dg(\hv) 
\end{aligned}
\label{equ:precision-est}
\end{equation}
where $\Sm$ is the sample covariance matrix, and $\Bm$ is the $n\times m$ incident matrix, specifying the connectivity between node pairs. 
The $m \times 1$ vector $\wv$ contains the weights associated with each link, and the $n \times 1$ vector $\vv$ are the self loops on each node. 

The above graph Laplacian learning algorithms are based on the assumption that 
all the entries of an observation, $\fv^i$, are available.
However, in an SAI without demosaicking, each pixel position contains at most one color component out of R, G and B. The observed block for each color is represented as a columnized vector
\begin{equation}
\fv^i = \begin{bmatrix}
\fv^i_\mathcal{O} \\
\fv^i_\mathcal{M} 
\end{bmatrix},
\end{equation}
where $\fv^i_\mathcal{O}$ is a $r$-dimensional vector containing the observed pixel intensities, and $\fv^i_\mathcal{M}$ of dimension $n-r$ represents the missing data. $r$ is a variable that changes between blocks. In order to optimize the graph connecting available pixels, we assume $\fv^i_\mathcal{O}$ to be a sub-sample of $\fv^i$ modeled as GMRF. 

In the statistics literature, many methods are proposed in estimating the inverse covariance matrix in a GMRF based on sub-sampled observations \cite{stadler2012missing,lounici2014high,kolar2012consistent}. In this work, we adopt the \textit{plug-in} algorithm by Kolar and Xing \cite{kolar2012consistent} due to its simplicity. 
The method consists of two steps. 
First, the sample covariance matrix is estimated using the incomplete observations. 
Given $f^i_a$ is the $a$-th variable in the observation $\fv^i$, we define $\rv^i$ as its indication vector, where element
\begin{equation}
\begin{cases}
   r^i_a = 1 \quad \text{if } f^i_a \text{ is available} \\
   r^i_a = 0 \quad \text{otherwise},
\end{cases}
\end{equation}
the estimated sample covariance matrix $\tilde{\Sm}$ is calculated as 
\begin{equation}
\tilde{S}_{a,b} = \frac{\sum_{i=1}^p r^i_a \cdot r^i_b \cdot (f^i_a-\mu^i_a)(f^i_b-\mu^i_b)}{\sum_{i=1:p}r^i_a \cdot r^i_b}
\end{equation}
See \cite{kolar2012consistent} for details and a theoretical justification of the estimation. 
In the second step, $\tilde{\Sm}$ will be \textit{plugged} into the objective function of the maximum likelihood estimation of the precision matrix $\Qm$. In this work, we optimize $\Qm$ based on Equ.\;\eqref{equ:precision-est} with $\Sm$ replaced by $\tilde{\Sm}$. 

Then, the precision matrix $\Qm$, and therefore the graph Laplacian matrix, can be derived for the whole block, including the pixel positions with missing data. We can decompose precision matrix $\Qm$ of the GMRF, which is the inverse of covariance matrix $\Sigmam^{-1}$, as 
\begin{equation}
\begin{aligned}
\Qm &= \begin{bmatrix}
\Qm_{O, O} & \Qm_{O,M} \\
\Qm_{M, O} & \Qm_{M,M}
\end{bmatrix},
\end{aligned}
\end{equation}
based on the subset of observed ($O$) and missing data ($M$).  

The inverse covariance matrix of available pixels, which defines the proximity between pixels in $\fv^i_\Oc$, can be calculated by taking the Schur complement of the sub-matrix $\Qm_{M,M}$ of $\Qm$:
\begin{equation}
\Lc = \Sigma_{O,O}^{-1} = \Qm_{O,O} - \Qm_{O,M}\Qm_{M,M}^{-1}\Qm_{M,O}.
\end{equation}

\begin{figure}[H]
\centering
\begin{subfigure}{\linewidth}
  \centering
  \includegraphics[width=0.8\linewidth]{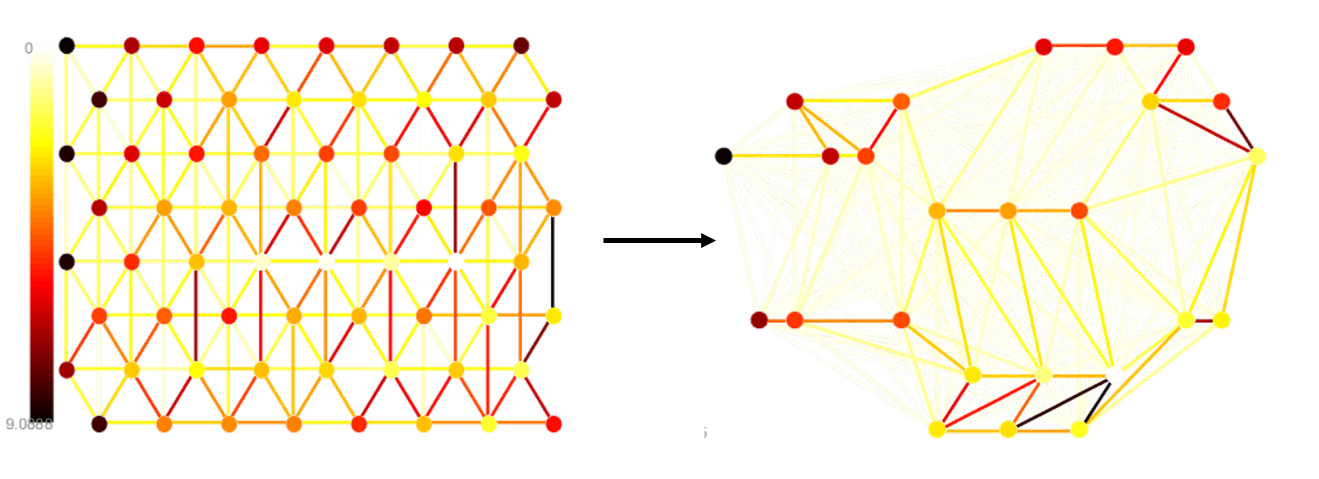}
  \caption{$\theta = \dfrac{3\pi}{8}$}
\end{subfigure}%
\hspace{1mm}
\begin{subfigure}{\linewidth}
  \centering
  \includegraphics[width=0.8\linewidth]{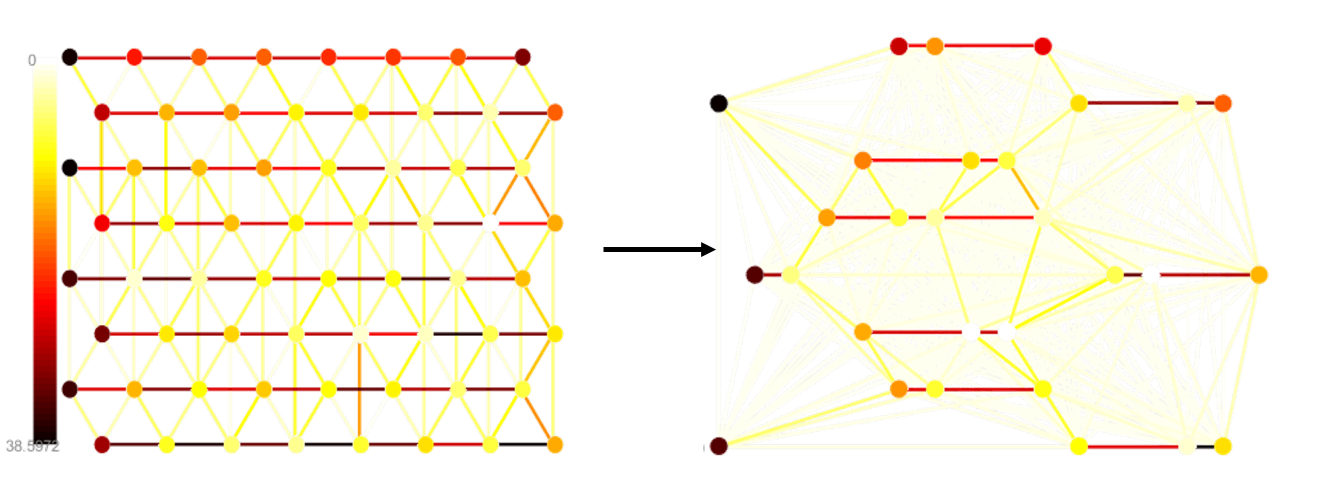}
  \caption{$\theta = 0$ (horizontal)}
\end{subfigure}
\vspace{-5mm}
\caption{Graph structure optimized for classes corresponding to intra-prediction angle (1) $\theta = \dfrac{3\pi}{8}$ (nearly vertical) and (2) $\theta = 0$ (horizontal). The link color indicates the associated weight and the node color indicates the associated self loop weight (darker: larger weight)}
\label{fig:illustrative-graph-opt}
\vspace{-5mm}
\end{figure}

In Fig.\;\ref{fig:illustrative-graph-opt}, two illustrative examples are shown for weighted graph optimized for set $\mathcal{M}_{\theta = \frac{3\pi}{8}}$ and  $\mathcal{M}_{\theta = 0}$. The figures in the left column are the regular graphs of the two classes learnt from the training set. 
The figures in the right column are examples of graphs connecting available sparsely distributed pixels, derived using the Schur complement and sparsification.
The nodes with strong self-loop weights concentrate around the boundaries close to the reference blocks, which have a better prediction. Therefore, the associated residuals on those pixels tend to have lower variance. 

\subsection{Graph-based Lifting Transform}

Due to the large data size of LF images, we apply the localized graph-based lifting transform (GLT), which was first proposed in \cite{narang2009lifting}. 
The GLT  is a multi-resolution, critically sampled filterbank. It consists of three building blocks: bipartition, prediction, and update. In bipartition, nodes are divided into two disjoint sets $\Uc$ and $\Pc$. Then, graph signal $\fv_\Uc$ attached to nodes in $\Uc$ is used to predict signal $\fv_\Pc$ in $\Pc$. The resulting prediction residuals are stored in $\Pc$ as the high frequency transformed coefficients $\cv_\text{H}$. In the update step, $\cv_\text{H}$ is used to update $\fv_\Uc$, which leads to a low frequency approximation $\cv_\text{L}$ of the input signal. The process can be written in matrix form as follows. 
\begin{align}
\text{Prediction:}\quad &\begin{bmatrix}
\cv_\text{H} \\
\fv_\Uc
\end{bmatrix} = \begin{bmatrix}
\Id & -\Pm \\
\zerov & \Id
\end{bmatrix}\begin{bmatrix}
\fv_\Pc \\
\fv_\Uc
\end{bmatrix} \\
\text{Update:} \quad &\begin{bmatrix}
\cv_\text{H} \\
\cv_\text{L}
\end{bmatrix} = \begin{bmatrix}
\Id & \zerov \\
\Um & \Id 
\end{bmatrix}\begin{bmatrix}
\cv_\text{H} \\
\fv_\Uc
\end{bmatrix},
\end{align}
where $\Pm$ and $\Um$ denote the linear operations for prediction and update,  respectively. 
The transform is guaranteed to be invertible regardless of how the nodes are bi-partitioned and how $\Pm$ and $\Um$ are designed. 

In the context of compression, the main objective in designing the lifting scheme is to steer most of the data energy to the low frequency band $\Uc$, namely to reduce the energy of prediction residuals $\cv_\text{H} = \fv_\Pc - \Pm\fv_\Uc$ in its complement $\Pc$. For signal $\fv$ modeled as GMRF $\Nc(\muv,\Qm^{-1})$, the minimum mean square estimator of signal value $f_i$ on node $v_i\in \Pc$ is expressed as
\begin{equation}
\begin{aligned}
\tilde{f}_i &= \mu_i - \sum_j\frac{Q_{i,j}}{Q_{i,i}}(f_j-\mu_j) \\
&= - \sum_j\frac{Q_{i,j}}{Q_{i,i}}f_j + (\mu_i + \sum_j\frac{Q_{i,j}}{Q_{i,i}} \mu_j).
\end{aligned}
\end{equation}
If the estimated precision matrix satisfies the graph Laplacian structure, as described in Equ.\;\eqref{equ:laplacian_struct} in the previous subsection, and if $\muv_i \approx \muv_j = \mu$, then the above equation can be simplified to
\begin{equation}
\begin{aligned}
\tilde{f}_i &= - \sum_j\frac{Q_{i,j}}{Q_{i,i}}f_j + \mu(1 + \sum_j\frac{Q_{i,j}}{Q_{i,i}}) \\
& = \frac{1}{D_{i,i} + H_{i,i}}\sum_j A_{i,j} f_j + \frac{H_{i,i}}{D_{i,i}+H_{i,i}}\mu.
\end{aligned}
\label{equ:prediction-with-self-loop}
\end{equation}
The estimator is simply a weighted average of the signal values on the neighboring nodes of $v_i$ and its associated mean. The self-loop weight $\Hm_{i,i}$ can be interpreted as a measurement of similarity to the mean $\mu$. In our experiment, we assume $\mu$ to be $0$ based on the observation that SAIs after intra-prediction mostly have low magnitude.
Note that if the graph is bipartite, \textit{i.e.} every node in $\Uc$ only has connections to nodes in $\Pc$, then the proposed filterbanks are optimal in terms of mean square error for GMRF. The prediction operation in Equ.\;\eqref{equ:prediction-with-self-loop}, when self loop weight $\Hm_{i,i}$ of each node is equal to $0$, is equivalent to the low complexity CDF53 filterbanks 
\begin{equation}
\tilde{f}_i = \frac{1}{D_{i,i}}\sum_{v_j\in \Uc} A_{i,j} f_j.
\end{equation}

For bipartition, we apply the algorithm proposed in \cite{nguyen2015downsampling} using maximum spanning tree to assign the available pixels to update and prediction set with linear time complexity. 
However, when graphs are not bipartite, the information of proximity on the links connecting nodes in the same set cannot be utilized, which leads to a suboptimal solution for prediction. 
Therefore, we apply the re-connection algorithm proposed in \cite{chao2016graph} to transform them into bipartite ones. The algorithm re-connects each node $v_i \in \Pc$ to be predicted to nodes in the opposite set $\Uc$ using Kron reduction \cite{dorfler2013kron}. 
Then, a simple sparsification is applied, which keeps only $k$ ($k=4$ in the experiment) links with the largest weights for each node in $\Pc$, in order to achieve localization and low complexity in prediction. 
The generalized CDF5/3 predictor in Equ.\;\eqref{equ:prediction-with-self-loop} is applied on the newly formed bipartite graph. 
It can be proven that without the sparsification, the applied predictor is equivalent to the minimum mean square error estimator (MMSE) for $\fv_\Pc$ given $\fv_\Uc$. 
Moreover, with the help of Kron reduction, which can be implemented iteratively, we can achieve lower complexity in implementation than directly computing the MMSE estimator, which requires computing matrix inverse.  

Same as the work in \cite{chao2016graph}, the design of the update filter is based on \cite{shen2009tree}, which promotes the orthogonality of the filterbanks, and the construction of graphs for lifting transform at levels $2$ and higher is based on Kron reduction. The transformed coefficients are uniformly quantized and reordered based on the approach in \cite{martinez2011lifting}. For entropy coding, we apply the Amplitude and Group Partitioning (AGP) method proposed by Said and Pearlman in \cite{said1997low}. 

\section{Experiments}
\label{Experiments}


\subsection{Dataset}

The LF images used in our experiments were from the EPFL Light-field data set in the JPEG Pleno Database \cite{LFdatabase}. 
The raw data were captured with a Lytro Illum camera \cite{rerabek2016new}. 
The LF data used for training and testing are listed in Table\;\ref{tab:dataset}. 
Four SAIs selected from each of the LF images in the training set were used for graph learning. 
Each test image was of size $5368\times7728$. 
In the baseline scheme, the raw data was converted into $15\times15$ full-color SAIs. 
Each SAI was of size $434\times 625$. We cropped the SAIs into $432\times 624$, so that the dimensions were multiples of the minimum coding unit of baseline schemes. 

\begin{table}[h]
\centering
\captionof{table}{LF data for training and testing}
\resizebox{.8\linewidth}{!}{%
  \begin{tabular}{ | c || c |}
    \hline
    \textbf{Training set} & \textbf{Test set} \\ \hline \hline
    Ankylosaurus\_\&\_Stegosaurus & Friends\_1 \\ \hline
    Ceiling\_Light & Bikes \\ \hline
    ISO\_Chart\_16 &  Flowers\\ \hline
    Perforated\_Metal\_1 & Ankylosaurur \& Diplodocus 1 \\ \hline
    Sophie\_\&\_Vincent\_3 &  \\ \hline
    Yan\_\&\_Krios\_standing &  \\ \hline
  \end{tabular}  }
  \label{tab:dataset}
\end{table}

\begin{figure}[H]
\begin{center}
\begin{subfigure}[t]{0.13\textwidth}
\includegraphics[width = \linewidth]{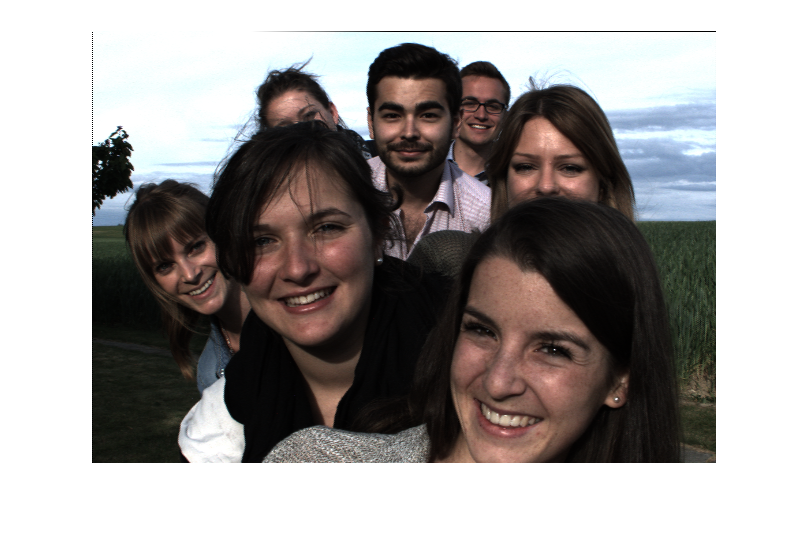}
\caption{}
\end{subfigure}
\hspace{-5mm}
\begin{subfigure}[t]{0.13\textwidth}
\includegraphics[width = \linewidth]{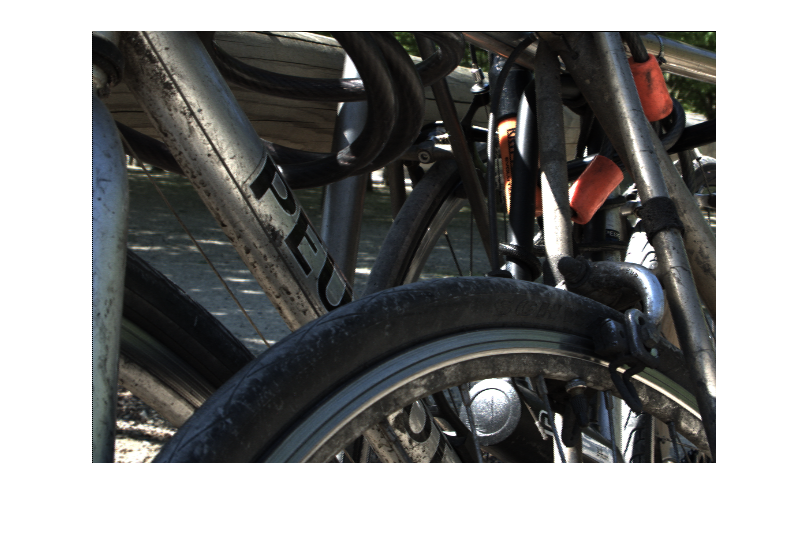}
\caption{}
\end{subfigure}
\hspace{-5mm}
\begin{subfigure}[t]{0.13\textwidth}
\includegraphics[width = \linewidth]{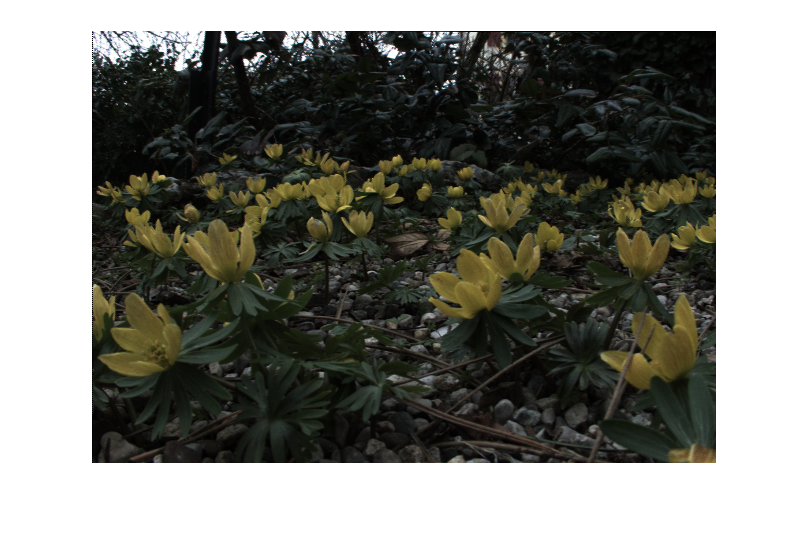}
\caption{}
\end{subfigure}
\hspace{-5mm}
\begin{subfigure}[t]{0.13\textwidth}
\includegraphics[width = \linewidth]{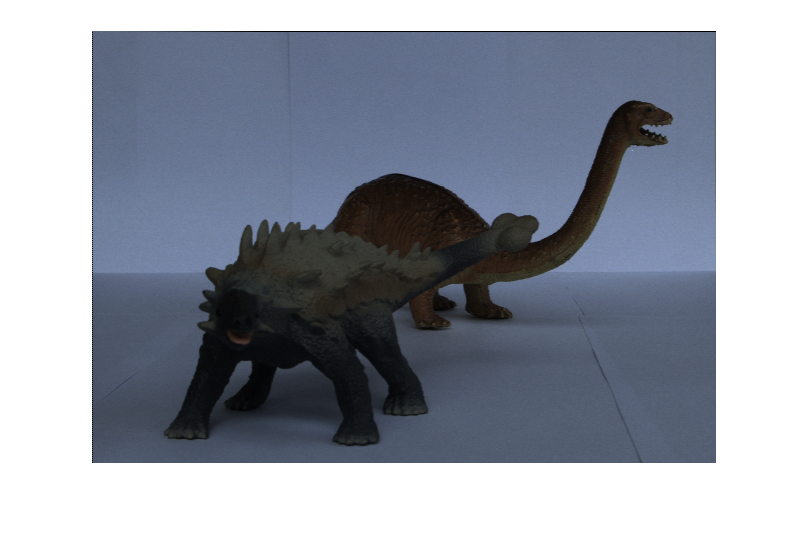}
\caption{}
\end{subfigure}
\end{center}
\vspace{-5mm}
\caption{The central view of the test images (a) Friends1 (b) Bikes, (c) Flowers, and (d) Ankylosaurur \& Diplodocus 1}
\label{fig:thumbnail-of-the-test-images}
\end{figure}


\subsection{Baseline Schemes}

We compare our proposed method with state-of-the-art coding solutions. 
The following anchor coding solutions were used as benchmarks, namely:
\begin{enumerate}
\item High Efficiency Video Coding all intra-prediction (HEVC-AI) / random access (HEVC-RA) / screen content coding extension (HEVC-SCC): 
HM-16.22 \cite{HM} developed by JCT-VC. 
\item Versatile Video Coding all intra-prediction (VVC-AI) / random access (VVC-RA): 
VTM-10.0 \cite{VTM} developed by JVET. 
\end{enumerate}
For fair comparison against our scheme, post filters, including deblocking filter, sample adaptive offset (SAO), and adaptive loop filter (ALF), were all disabled in the baseline. 




\subsection{Experimental Setting}
\subsubsection{First series of experiments}
\label{sec:1st-experiment}
We consider three different scenarios for the graph-based coding scheme: 
\begin{enumerate}
\item DGLT: geometric distance based graph lifting transform without intra-prediction, proposed in \cite{chao2017pre}.
\item intraDGLT: geometric distance based graph lifting transform with the proposed intra-prediction (Section \ref{prediction}).
\item intraLGLT: graph learning based graph lifting transform (Section \ref{sec:graph-learning}) with the proposed intra-prediction. 
\end{enumerate}

Since our scheme is a pre-demosaic coding technique, as a pre-demosaic coding baseline we chose HEVC encoding of the pre-demosaic lenselet images in 4:0:0 RGB. 
For extensive evaluation, we also encoded demosaicked lenselet images with HEVC-AI and VVC-AI in 4:4:4 RGB. 
Since the IBC technique in HEVC-SCC can further improve coding gain over HEVC, the HEVC-SCC encoding of the full-color lenselet images in 4:4:4 RGB was also compared.

\subsubsection{Second series of experiments}
\label{sec:2nd-experiment}
$15\times15$ SAIs were encoded as a pseudo video sequence created using a serpentine scanning order. 
The order is defined in JPEG Pleno Light Field Coding Common Test Condition \cite{CTC}. 
In addition to exploiting both intra- and inter-view redundancy using HEVC-RA and VVC-RA configurations, since we are focusing on coding schemes that allow efficient random access, the baselines coded with All-Intra configuration were also included, where every SAI was coded as a single 2D image with HEVC-AI and VVC-AI in original 4:4:4 RGB and 4:2:0 YUV formats.
In this part, only the best graph-based coding scheme, intraLGLT, was used to compare against the above-mentioned anchor coding solutions.

\subsection{Evaluation Metrics}

The evaluation metrics were the number of bits per pixel (bpp) and average peak signal-to-noise ratio (PSNR) over 4:4:4 RGB format of reconstructed SAIs.

For archival purposes, it is more appropriate to evaluate the quality of reconstruction in RGB color space and on the reconstructed lenselet images, \textit{i.e.}, all captured pixels should be reconstructed. 
However, current state-of-the-art schemes using HEVC discard under-exposed pixels at the boundary of macro-pixels during the conversion to SAI array.
Moreover, the interpolation involved in the resampling process 
is irreversible, so that we cannot recover the original lenselet image from the SAI. 
Hence, for evaluation, performances were compared against the full-color SAIs before compression (ground truth), generated from the raw lenselet data using the demosaicking and calibration pipeline described in \cite{dansereau2013decoding, LFtoolbox}.  

For the 4:2:0 YUV format, the reconstructed SAIs were translated back to 4:4:4 RGB format before evaluation. 
The up-sampling for U and V components was based on nearest neighbor interpolation. 
For HEVC/VVC encoding of the demosaicked lenselet image in Section \ref{sec:1st-experiment}, the reconstructed lenselet image was decomposed into SAIs for comparison.

\subsection{Parameter Setting}

\textbf{Proposed.} Each SAI was divided into non-overlapped $8 \times 8$ blocks. 
The parameter $\delta$ in Equ.\;\eqref{equ:tenstor-estimation-weights} was set to $0.9$, and the parameters $\sigma, p_1, p_2$ in Equ. \eqref{equ:adaptive-Gaussian-kernal} and \eqref{equ:weight-numerical-stable} for data-adaptive Gaussian kernal were chosen to be $1.6$, $0.001$ and $0.001$, respectively. 
The threshold value $T$ in Equ.\;\eqref{equ:training-classification} for training block classification was set to $1.5$. 
We applied $2$-level lifting transform for transform coding for pixels in each block. 
The selected QP values ranged from $4$ to $36$.

\textbf{Anchor.} The profile, bit depth, GoP and QP settings for the anchor coding schemes in the experiments are summarized in Table\;\ref{tab:experiment-setting}.
\begin{table}[H]
  \centering
\begin{tabular}{ |p{4.5cm}||p{0.5cm}|| p{1cm} | p{1cm}| }
\hline
Profile: main\_444\_RExt & GoP & \multicolumn{2}{c}{QP (start:stride:end)} \vline \\
Bit depth: 8-bit & & YUV420 & RGB444 \\
 \hline
The 1st series of experiments & & &\\
 \hline
HEVC-SCC All Intra-prediction (lenselet) & 1 &   & 11:3:35 \\
HEVC All Intra-prediction (predemosaic lenselet) & 1 &   & 2:3:35 \\
HEVC All Intra-prediction (lenselet) & 1 &   & 11:3:35 \\
VVC All Intra-prediction (lenselet) & 1 &   & 11:3:35 \\
 \hline
The 2nd series of experiments & & &\\
 \hline
 HEVC All Intra-prediction  & 1 &5:3:29 & 14:3:35 \\
VVC All Intra-prediction & 1 &  & 14:3:35 \\
HEVC Random Access & 8 &  & 5:3:29 \\
HEVC Random Access & 16 & 2:3:26 & 5:3:29 \\
VVC Random Access & 16 &  & 5:3:29 \\
 \hline
\end{tabular}
 \caption{Summary of parameter settings of anchor coding solutions in the experiments} 
 \label{tab:experiment-setting}
\end{table}

\begin{figure*}[t!]
\begin{center}
\begin{subfigure}[t]{0.39\textwidth}
\includegraphics[width = \linewidth]{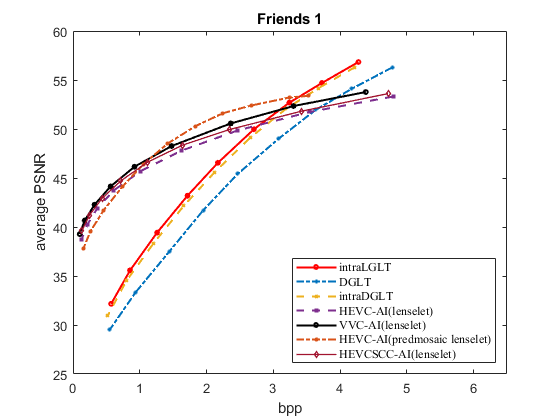}
\vspace{-6mm}
\caption{}
\end{subfigure}
\hspace{-6mm}
\vspace{-2mm}
\begin{subfigure}[t]{0.39\textwidth}
\includegraphics[width = \linewidth]{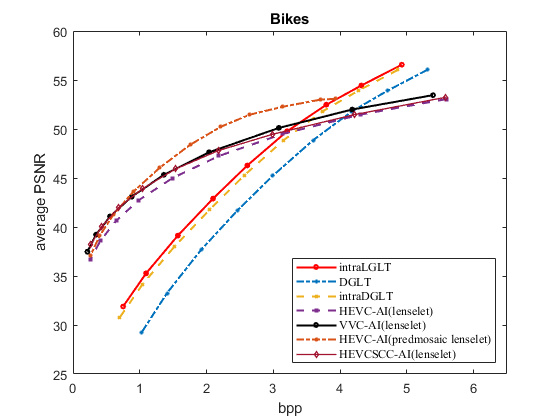}
\vspace{-6mm}
\caption{}
\end{subfigure}
\begin{subfigure}[t]{0.39\textwidth}
\includegraphics[width = \linewidth]{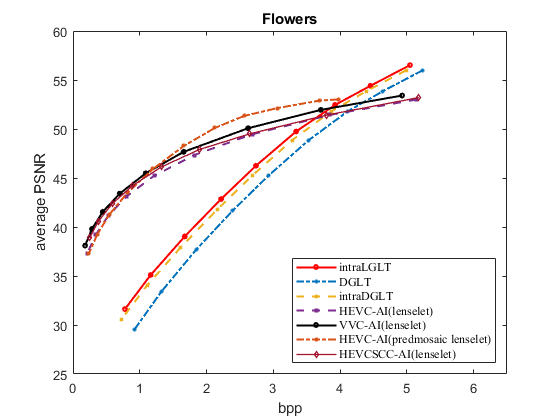}
\vspace{-6mm}
\caption{}
\end{subfigure}
\hspace{-6mm}
\vspace{-2mm}
\begin{subfigure}[t]{0.39\textwidth}
\includegraphics[width = \linewidth]{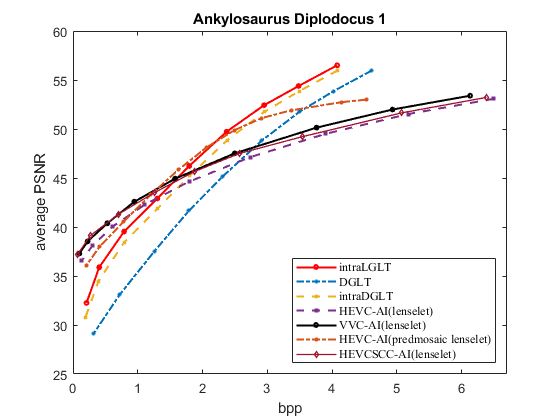}
\vspace{-6mm}
\caption{}
\end{subfigure}
\end{center}
\vspace{-5mm}
\caption{PSNR-RGB performance in the first series of experiments: (a) Friends1 (b) Bikes, (c) Flowers, and (d) Ankylosaurur \& Diplodocus}
\label{fig:1st_curve}
\vspace{-2mm}
\end{figure*}

\begin{figure*}[t!]
\begin{center}
\begin{subfigure}[t]{0.39\textwidth}
\includegraphics[width = \linewidth]{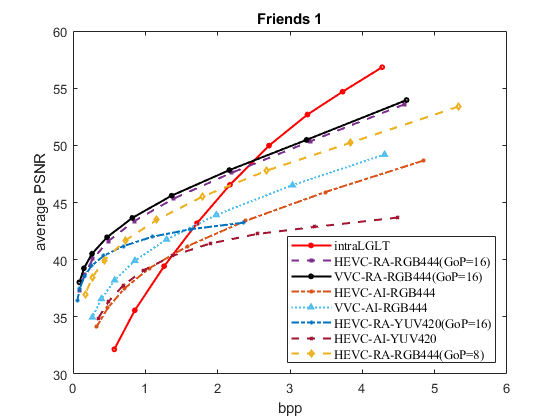}
\vspace{-6mm}
\caption{}
\label{fig:psnr_friends}
\end{subfigure}
\hspace{-6mm}
\vspace{-2mm}
\begin{subfigure}[t]{0.39\textwidth}
\includegraphics[width = \linewidth]{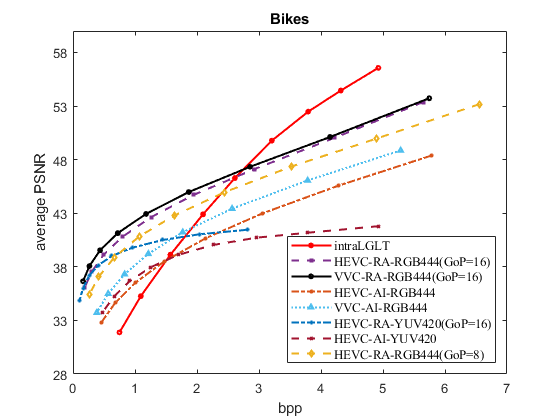}
\vspace{-6mm}
\caption{}
\label{fig:psnr_Bikes}
\end{subfigure}
\begin{subfigure}[t]{0.39\textwidth}
\includegraphics[width = \linewidth]{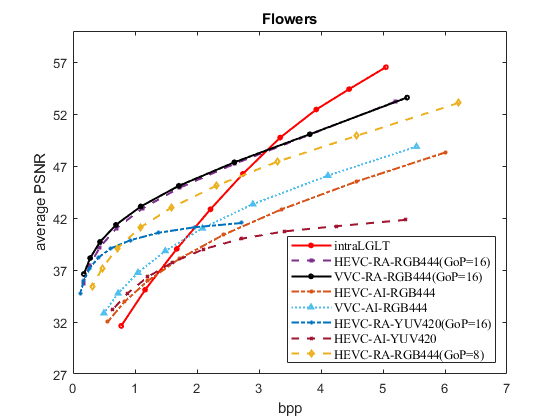}
\vspace{-6mm}
\caption{}
\label{fig:psnr_flowers}
\end{subfigure}
\hspace{-6mm}
\vspace{-2mm}
\begin{subfigure}[t]{0.39\textwidth}
\includegraphics[width = \linewidth]{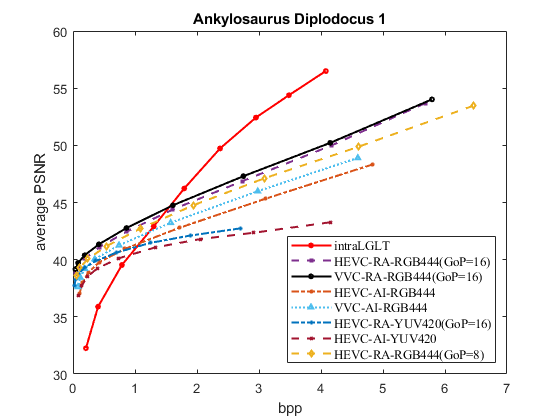}
\vspace{-6mm}
\caption{}
\label{fig:psnr_bikes}
\end{subfigure}
\end{center}
\vspace{-4mm}
\caption{PSNR-RGB performance in the second series of experiments (a) Friends1 (b) Bikes, (c) Flowers, and (d) Ankylosaurur \& Diplodocus}
\label{fig:2nd_curve}
\vspace{-2mm}
\end{figure*}

\subsection{Experimental Results and Analysis}

\subsubsection{The first series of experiments} Fig.\;\ref{fig:1st_curve} shows the PSNR comparisons for test images in Table \ref{tab:dataset}.

For our scheme, with the proposed intra-prediction, the correlation between neighboring blocks were exploited to reduce redundancies before compression, leading to 3dB performance gain over the setup where no intra-prediction was performed \cite{chao2017pre}. 
Moreover, with graph learning, local directional characteristics were exploited to determine graph connections, which provided better accuracy in modeling similarity among pixels compared to connections based solely on Euclidean distance. 
We observe that, using graph learning, around $0.5$dB gain was achieved against a graph design based solely on distance when intra-prediction was used.

Compared with other anchor schemes, the HEVC encoding of the pre-demosaic lenselet images has significant coding gain at high bitrates. However, at lower rates, its performance dropped more drastically compared with the HEVC encoding of demosaicked lenselet, which will be discussed in more detail in Section.\;\ref{sec:analysis}. With more advanced intra-prediction techniques, VVC-AI outperformed HEVC-AI at any rate with $1dB$ performance gain. HEVC-SCC outperformed HEVC by using additional tools, e.g., IBC, to exploit similarity between neighboring MIs in prediction.

For applications such as archival and instant storage on cameras, images are stored in very high quality. Therefore, in the evaluation, we focused mainly on the high rate region. 
For bpp $>$ 3, graph-based coding schemes outperformed baseline schemes applying HEVC/VVC in demosaicked lenselet image.  Applying HEVC/VVC intra coding on the whole lenselet image provided significant gains at low bit rate over the proposed scheme, which can be caused by the lack of accuracy in the proposed structure tensor based intra-prediction. The estimation of structure tensor in the target block depends on the gradients of the reconstructed reference blocks. When reference blocks are coded at low bit rate, the gradients may fail to reflect the characteristics, \textit{e.g.}, edge direction, in the target block. Although HEVC/VVC intra coding on the whole lenselet image outperforms the proposed scheme at low bit rate, it introduces large dependencies among SAIs, which results in a less efficient random access.

\subsubsection{The second series of experiments} 
\label{sec:2nd-analysis} 

Fig.\;\ref{fig:2nd_curve} shows that, for higher bitrate (bpp $>$ 2), graph-based coding schemes significantly outperformed the baseline schemes applying HEVC-AI and VVC-AI to demosaicked SAIs. HEVC-RA and VVC-RA had similar performance and outperformed HEVC/VVC-AI, because there exist only small disparity, and therefore high correlation, between SAIs caused by light rays coming from different directions. 
Further, the anchor coding schemes of GoP equal to $16$ outperformed those of GoP equal to 8 by using more inter-prediction between frames while sacrificing random access efficiency. 
Although the proposed LGLT scheme performed worse than the anchors at low bitrates, it achieved much higher coding efficiency at high bitrates, making it attractive for archival applications.

To substantiate the claim that, for the archiving purpose, images are stored at high bitrates, we encoded SAIs of the four test images with the default JPEG setting that the iPhone applies on captured images, and measured their individual bitrates. 
The archival bitrate range measured is summarized in Table\;\ref{tab:archive}. 
As observed in Fig.\;\ref{fig:2nd_curve}, in the archival bitrate range identified, our scheme outperformed all the All-Intra coding schemes, with around 3dB performance gain over VVC-AI and 5dB performance gain over HEVC-AI. 
Compared to the baseline schemes using inter-prediction, our scheme has the advantage of random access, which is analyzed in Section\;\ref{sec:Complexity-Analysis}.
Next, we provide insights into why our method has advantages at very high rates.

\begin{table}[h]
\centering
\resizebox{.6\linewidth}{!}{%
  \begin{tabular}{ | c || c |}
    \hline
    \textbf{Test images} & \textbf{Bitrate (bpp)} \\ \hline \hline
    Friends\_1 & 2.05 - 2.24\\ \hline
    Bikes & 2.45 - 2.66 \\\hline
    Flowers & 2.71 - 2.93 \\\hline
    Ankylosaurur \& Diplodocus 1 & 1.45 - 1.61\\ \hline
  \end{tabular}  }
  \caption{ The bitrates of SAIs encoded by JPEG with default setting from the iPhone}
  \label{tab:archive}
\end{table}

\subsubsection{ Advantages of proposed method at high bitrates}
\label{sec:analysis}
{In a conventional scheme, the dimension of raw images to be encoded is increased to obtain full-color images via interpolation and demosaicking. 
This is a minor problem at low rates, since the interpolated images are smooth and can be interpolated with few transform coefficients. 
However, as the bitrate increases, higher frequency coefficients are no longer quantized to zero, and encoding more coefficients increases the rate substantially. 
Instead, in our proposed scheme, demosaicking and interpolation are postponed, so that the total number of coefficients to be coded is smaller, resulting in substantially better performance at high rates, which also explains why the HEVC encoding of pre-demosaic lenselet images can significantly outperform the other anchors in the high-rate region.


Moreover, because compression is performed prior to interpolation in our proposed method, 
compression noise in encoded pixels is spread to interpolated pixels. 
Thus, at low rates, error in encoded pixels, which is high, is propagated via interpolation, resulting in worse performance than conventional methods.
Instead, at high rates, since the number of pixels to be encoded is reduced, per pixel distortion is significantly lower than the conventional method, and the effect of error propagation via interpolation is negligible. Thus, at high rates, the pre-demosaic coding technique can outperform the conventional techniques applied to demosaicked images. 

In comparison to the HEVC coding of pre-demosaic lenslet images, the graph structure provides better modeling of pixel similarities. Specifically, a graph connects a pixel only to its adjacent pixels of the same color, avoiding the discontinuities between different color components, and thus achieves better coding efficiency at high bit rates against coding the Bayor patterned pre-demosaic lenslet images.

}

\begin{table}[t]
  \centering
\begin{tabular}{ |p{1.8cm}||p{1.8cm}|| p{1.8cm} | p{1.8cm}| }
\hline
&  HEVC based & \multicolumn{2}{c}{Proposed scheme} \vline \\
 \hline
 & & w/o Parallelization & Parallelization \\
 \hline
 Intra-prediction  & \textit{O($N$)} & \textit{O($N^2$)} & \textit{O($N$)} \\
 \hline
 Transform & \textit{O($N\log N$)} &  \textit{O($N$)}  & \textit{O($N$)} \\
 \hline
\end{tabular}
 \caption{ Time complexity analysis of HEVC based coding scheme and the proposed graph based coding scheme} \label{tab:complexity}
\end{table}

\subsection{Complexity Analysis}
\label{sec:Complexity-Analysis}
1. \textit{Time Complexity Analysis}

Time complexity of our proposed coding scheme is summarized in Table\;\ref{tab:complexity} for the cases with and without parallelization, compared to the complexity in HEVC based coding scheme. 
In the HEVC/VVC coding scheme, intra-prediction of each pixel in the prediction unit is computed as the weighted average of up to two reference samples in the block boundary, and thus time complexity $\textit{O}(N)$ is needed to predict \textit{N} data points. 
The DCT2 transform is applied horizontally/vertically, leading to time complexity $\textit{O}(N\log N)$. Note that in HEVC/VVC, the intra-prediction mode and the size of a coding unit are determined by recursively splitting a coding tree unit and comparing rate-distortion cost, which potentially increases the complexity in comparison to the proposed intra-prediction in the graph-based methods. 

In our proposed intra-prediction, the data-adaptive steering kernel is used to calculate weights for reference pixels. 
In the worst case, one pixel in the predicted block is predicted with the weighted average of reference pixels in the four neighboring blocks, resulting in $\textit{O}(N^2)$ to predict $N$ data points. 
If the weight of each reference sample is calculated in parallel, the time complexity for intra-prediction can be reduced to $\textit{O}(N)$. 
The graph-based lifting transform is applied on the intra-predicted residuals. Note that the pattern of available pixels on each SAI changes depending on the camera settings, and hence different graphs need to be constructed for different cameras. 
Each pixel (node) is connected to its $4$-nearest neighbors in our experiment, which requires $\textit{O}(N)$ in computation.
The prediction and update filters---CDF 5/3 filterbanks---used in the proposed scheme are localized, \textit{i.e.}, computation of the transform coefficient of each pixel involves only its connected neighbors on the bipartite graph. 
In addition, only $2$-level lifting transform is applied. 
Therefore, for $N$ data points, the complexity of transform in our proposed coding scheme is $\textit{O}(N)$. 
As a result, comparable decoding complexity can be achieved at decoder for our proposed coding scheme compared to HEVC based coding scheme if parallelization is applied.

2. \textit{Random Access Complexity Analysis}

{We define random access complexity as the number of reference frames needed for decoding to access a chosen view. 
We consider the average and the maximum number of reference frames required to be decoded for a view as the average and worst-case random access complexity respectively. We evaluate the random access complexity of the coding schemes in the experiments by using the following criteria \cite{Mehajabin2019}:
\begin{equation}
    C_{\text{average}}^{\text{RA}} = \frac{1}{N}\sum_{x=1}^{N}f(x)
\end{equation}
\begin{equation}
    C_{\text{worst-case}}^{\text{RA}} = \max_{x=1...N}{f(x)}
\end{equation}
where $N$ is the total number of views, and $f(x)$ is a function returning the number of SAIs required to be decoded for accessing the view $x$. 
The results of random access complexity are summarized in Table \ref{tab:RAcomplexity}.

\begin{table}[h]
\centering
\resizebox{.8\linewidth}{!}{
\begin{tabular}{ |p{3.5cm} |p{1.5cm}|p{1.5cm} |}
\hline
 \textbf{\footnotesize Method} & \textit{\footnotesize  Worst-case} & \textit{\footnotesize  Average} \\
 \hline
 { \footnotesize HEVC/VVC-AI} &
 { \footnotesize $0$} & { \footnotesize $0$} \\
  \hline
 { \footnotesize HEVC/VVC-RA (GoP=8)} &
 { \footnotesize $4$  } & { \footnotesize $2.98$  } \\
 \hline
  { \footnotesize HEVC/VVC-RA (GoP=16)} &
   { \footnotesize $5$  }  &  { \footnotesize $3.98$  } \\
 \hline
   { \footnotesize Proposed scheme} &
 { \footnotesize $0$} & { \footnotesize $0$} \\
 \hline
\end{tabular}}
 \caption{Random access complexity analysis of anchor coding solutions in RA and AI coding scheme and the proposed graph based coding scheme} \label{tab:RAcomplexity}
\end{table}

Although applying HEVC/VVC intra coding on the whole lenselet image and inter coding on the PVS-based SAIs provides significant gains at low bitrates over schemes coding each SAI separately, it also introduces large dependencies among SAIs. 
That is, in order to access a specific SAI, the user has to decode the whole lenselet image, which greatly increases latency and memory buffer requirements. 
In schemes where the SAIs are coded separately, on the other hand, an SAI can be encoded/decoded independently from each other. 

For evaluation purpose, we employed the demosaicking algorithm in \cite{malvar2004high} for both test and anchors. Note that while a specific demosaicking algorithm is employed in the experiment, the proposed scheme is targeting coding the pre-demosaic pixels, and is not explicitly tied to one particular demosaicking algorithm. The algorithm in the experiment requires a few neighboring pixels in up to $10$ reference SAIs to demosaick a single position.
In the literature, however, there exist many demosaicking methods that perform interpolation using far fewer neighboring color components, as described in \cite{niu2018low}, resulting in fast implementations and less dependency in our LF coding scheme. 
Tailoring for our coding scheme, a demosaicking algorithm that interpolates a missing color component using only available color pixels in the same SAI, such as the methods in \cite{xu2014multi} and \cite{seifi2014disparity}, can also be chosen to avoid introducing any dependency among SAIs on the decoder side, though they incur some marginal quality degradation.

Since intra-prediction is also a local operation, the demosaicking algorithm in the experiment can be implemented as a small moving window, e.g., $ 24 \times 24$ block, of local blocks in the reference SAIs that includes both the predictor and the target block containing reference pixels to a demosaicked pixel.
Although the process in the experiment entails some dependencies between an SAI and its neighbors, since pixels in each neighboring SAI can be decoded in parallel, 
the latency requirement is significantly lower than coding the whole lenselet image or the latency in the PVS-based HEVC scheme \cite{dai2015lenselet,liu2016pseudo, zhao2016light, hariharan2017low,vieira2015data, li2017pseudo}. In the worst case, 4 SAIs have to be decoded sequentially before accessing an SAI if the GoP is equal to 8, and 5 SAIs are required if the GoP is equal to 16.}

\section{Conclusion}
\label{conclusion}
In this paper, we describe a novel coding scheme for light field image based on the graph based lifting transform. 
Our scheme can encode the original raw data without introducing redundancies (due to demosaicking and calibration) or distortion (from color conversion and downsampling). Moreover, we propose an intra-prediction scheme and a graph learning algorithm for pixels in SAIs that are sparsely distributed. The pixels are then connected as graphs and encoded with low-complexity graph-based lifting transform. 
Coding results in the high rate region demonstrate that our proposed method outperforms the widely applied HEVC-based and the cutting-edge VVC-based approaches in All-Intra and random access configuration.


%

%

%
%

\ifCLASSOPTIONcaptionsoff
  \newpage
\fi



%
\bibliographystyle{ieeetr}
\bibliography{refs.bib}

\end{document}